\documentclass[a4paper,fleqn,usenatbib]{mnras}
\usepackage[T1]{fontenc}
\usepackage{ae,aecompl}
\usepackage{graphicx}	% Including figure files
\usepackage{amsmath}	% Advanced maths commands
\usepackage{amssymb}	% Extra maths symbols
\usepackage{makecell}
\usepackage{multirow}
\usepackage{ulem}
\usepackage{xcolor}

\newcommand{\kms}{km s$^{-1}$}
\newcommand{\dego}{$^\circ$}

\title[Narrow polar jets in the nascent wind of EP Aqr]{Observation of narrow polar jets 
in the nascent wind of oxygen-rich AGB star EP Aqr}
\author[P. Tuan-Anh et al.]{{P. Tuan-Anh$^1$\thanks{E-mail: ptanh@vnsc.org.vn}, D.T. Hoai$^{1,2}$, P.T. Nhung$^1$, P. Darriulat$^1$, P.N. Diep$^{1,2}$,}
\newauthor{T. Le Bertre$^3$, N.T. Phuong$^{1,2}$, T.T. Thai$^1$, J.M. Winters$^4$} 
\\
$^1$Department of Astrophysics, Vietnam National Space Center (VNSC), Vietnam Academy of Science and Technology (VAST),\\
18 Hoang Quoc Viet, Cau Giay, Ha Noi, Viet Nam\\
$^2$Graduate University of Science and Technology (GUST), Vietnam Academy of Science and Technology (VAST), \\
18 Hoang Quoc Viet, Cau Giay, Ha Noi, Vietnam\\
$^3$LERMA, UMR 8112, CNRS and Observatoire de Paris, PSL Research University, 61 av. de l'Observatoire, F-75014 Paris, France\\
$^4$IRAM, 300 rue de la Piscine, Domaine Universitaire, F-38406 St. Martin d'H\`{e}res, France\\
}
\date{Accepted XXX. Received YYY; in original form ZZZ}

\pubyear{2019}
\begin{document}
\label{firstpage}
\pagerange{\pageref{firstpage}--\pageref{lastpage}}
\maketitle

\begin{abstract}
  Using ALMA observations of $^{12}$CO(2-1), $^{28}$SiO(5-4) and $^{32}$SO$_2$(16$_{6,10}$-17$_{5,13}$) emissions of the circumstellar envelope of AGB star EP Aqr, we describe the morpho-kinematics governing the nascent wind. Main results are: 1) Two narrow polar structures, referred to as jets, launched from less than 25 au away from the star, build up between $\sim$ 20 au and $\sim$ 100 au to a velocity of $\sim$ 20 \kms. They fade away at larger distances and are barely visible in CO data. 2) SO$_2$, SiO and CO emissions explore radial ranges reaching respectively $\sim$30 au, 250 au and 1000 au from the star, preventing the jets to be detected in SO$_2$ data. 3) Close to the star photosphere, rotation (undetected in SiO and CO data) and isotropic radial expansion combine with probable turbulence to produce a broad SO$_2$ line profile ($\sim$ 7.5 \kms\ FWHM). 4) A same axis serves as axis of rotation close to the star, as jet axis and as axi-symmetry axis at large distances. 5) A radial wind builds up at distances up to $\sim$ 300 au from the star, with larger velocity near polar than equatorial latitudes. 6) A sharp depletion of SiO and CO emissions, starting near the star, rapidly broadens to cover the whole blue-western quadrant, introducing important asymmetry in the CO and particularly SiO observations. 7) The $^{12}$C/$^{13}$C abundance ratio is measured as 9$\pm$2. 8) Plausible interpretations are discussed, in particular assuming the presence of a companion.

\end{abstract}

%%%%%%%%%%%%%%%%%%%%%%%%%%%%%%%%%%%%%%%%%%%%%%%%%%

%%%%%%%%%%%%%%%%% BODY OF PAPER %%%%%%%%%%%%%%%%%%

\section{Introduction}  \label{sec1}

In a recent paper \citep{Hoai2019} we presented a detailed analysis of the circumstellar envelope of EP Aqr, an oxygen-rich M-type star on the early part of the AGB, at a distance of 114$\pm$8 pc from the Sun \citep{vanLeeuwen2007, Gaia2018}, based on ALMA observations of \mbox{$^{12}$CO(1-0)} and \mbox{$^{12}$CO(2-1)} emissions. It gives evidence for approximate axi-symmetry about an axis making an angle $\varphi \sim$ 10\dego\ with the line of sight and projecting on the sky plane some 20\dego\ west of north. The Doppler velocity spectra are shown to consist of a narrow central component associated with an equatorial outflow expanding at low velocity $V_{eq}\sim(0.33\pm0.03)/\sin\varphi$ \kms\ and a broad component associated with a bipolar outflow expanding at a velocity increasing with stellar latitude $\alpha$ up to 10 to 12 \kms\ near the poles. An upper limit of about a third of the expansion velocity is placed on a possible rotation velocity of the equatorial outflow. Both components are found to merge at an angular distance of $\sim$ 2 arcsec from the star ($\sim$ 200-250 au), below which the details of the morpho-kinematics could not be reliably evaluated. A joint description taking temperature and absorption effects in due account, gives evidence for an approximately constant mass loss rate of ($1.6\pm0.4$) $10^{-7}$ M$_{\odot}$year$^{-1}$, the flux of matter being maximal at intermediate stellar latitudes, and for a temperature decreasing with the distance $r$ from the star as $\sim 109$ [K] $\exp(-r$[arcsec]/3.1). The decrease of intensity due to self-absorption of the emitted radiation, evaluated from a radiative transfer calculation, is found at the level of 20\% on average, not exceeding $\sim$ 40\%. Irregularities of the emission, in both intensity and redshift, were described and possible interpretations were discussed. The analysis was compared with that of different but similar observations recently presented by \citet{Homan2018}.

In the present article, we extend the analysis to short distances from the star using new ALMA observations of SiO(5-4) and SO$_2$($16_{6,10}-17_{5,13}$) emissions (hereafter referred to simply as SiO and SO$_2$), probing the region where the wind builds up and where spherical symmetry is being broken. We compare the results with $^{12}$CO(2-1) emission. Contrary to the preceding analysis \citep[]{Hoai2019}, where a stationary axi-symmetric radial wind of constant velocity could give a fair description of observations, making it possible to de-project the effective emissivity with reasonable confidence, such simplifying hypotheses are no longer tenable at shorter distances from the star: the under-constrained nature of radio astronomy observations (two components of space and one of velocity being measured out of three of each) prevents a reliable interpretation to be proposed with certainty. In particular, local thermal equilibrium can no longer be taken as granted \citep[]{Freytag2017}; the wind can no longer be assumed to be radial \citep[]{Dorfi1996}, evidence for significant rotation being given by the SO$_2$ observations; it can no longer be assumed to be stationary, star pulsations being likely to play an important role in the mass loss mechanism of M stars \citep{McDonald2016}; and, obviously, it can no longer be assumed to be independent of the distance to the star. A cautious attitude is a must under such circumstances when proposing a plausible interpretation of the observations in terms of physical mechanisms governing the mass loss.  
After a brief description of the observations and of the reduction of the data, we review successively the SO$_2$, SiO and CO observations and analyse the properties of the morpho-kinematics of the circumstellar envelope that they imply. We conclude by attempting a unified description of the mass loss mechanism. We refer the reader to \citet{Hoai2019} for a detailed presentation of earlier studies.

\section{Observations and data reduction}  \label{sec2}
The observations used in the present article were made in Cycle 4 of ALMA operation (2016.1.00026.S) between 2016 October 30 and 2017 April 5. Details of the $^{12}$CO(2-1) observations, here referred to simply as CO, and of the associated data reduction were presented in \citet{Hoai2019} and are not repeated here. Together with observations of the CO line in ALMA Band 6, observations of $^{13}$CO(2-1) and SiO(5-4) line emissions were obtained. Original spectral setups, approximately centred on the lines, include 1920 channels $\sim$ 30.5 kHz wide for $^{13}$CO(2-1) and 960 channels $\sim$ 61 kHz wide for SiO(5-4). They were observed in two execution blocks in mosaic mode (10 pointings) with the number of antennas varying between 38 and 40. Both lines were observed in two different configurations, C40-2 and C40-5. 

We also use SO$_2$($16_{6,10}-17_{5,13}$) observations from another ALMA Cycle 4 project (2016.1.00057.S), made between 2016 October 8 and December 17. The spectral setup included 960 channels $\sim$ 488 kHz wide and approximately centred on the line. These observations were also made in two execution blocks in mosaic mode (3 pointings), in two configurations, with the number of antennas varying between 39 and 43. 

The data have been calibrated using CASA\footnote{http//casa.nrao.edu} standard scripts. After having obtained calibrated visibilities, we converted them to the Local Standard of Rest (LSR) kinematics spectral frame. Data were then smoothed to obtain a spectral resolution of 0.2 \kms\ for both $^{13}$CO(2-1) and SiO(5-4) observations, and 0.7 \kms\ for SO$_2$. Doppler velocities, measured with respect to the star systemic velocity, cover between $-$20 and +20 \kms\ for $^{13}$CO(2-1) and SO$_2$ and between $-$30 and +30 \kms\ for SiO. In all cases, merging was done in the \textit{uv} plane using as origin of the Doppler velocity scale the star LSR velocity of $-$33.6 \kms. Mapping was done with either GILDAS\footnote{https://www.iram.fr/IRAMFR/GILDAS} ($^{12}$CO(2-1), $^{13}$CO(2-1)) or CASA (SiO(5-4), SO$_2$($16_{6,10}-17_{5,13}$)). Natural weighting was used for both $^{12}$CO and $^{13}$CO lines after having merged C40-2 and C40-5 observations; but before merging more weight was given to the $^{13}$CO C40-5 observations to account for their weaker signal in comparison with C40-2 observations. The main parameters attached to these observations are listed in Table \ref{tab1}.

\begin{table*}
  \centering
  \caption{Main observation parameters.}
  \label{tab1}
  \begin{tabular}{ccccc}
    \hline
    &$^{12}$CO(2-1)&$^{13}$CO(2-1)&SiO(5-4)&SO$_2(16_{6,10}-17_{5,13}$)\\
    \hline
    Beam FWHM (arcsec$^2$)
    &0.33$\times$0.30
    &0.27$\times$0.25
&0.29$\times$0.25
&0.18$\times$0.17\\
Beam PA (degree)
&$-$80
&111
&$-$85
&$-$14\\
Maximum baseline (m)
&1400
&1400
&1400
&3144\\
Minimum baseline (m)
&15
&15
&15
&15\\
Time on source (min)
&52
&50
&50
&160\\
Mosaic coverage (arcsec$^2$)
&$\sim$60$\times$45
&$\sim$60$\times$45
&$\sim$60$\times$45
&$\sim$17$\times$15\\
Noise (mJy beam$^{-1}$channel$^{-1}$)&6&3.8&3.9&1.0\\
\hline
  \end{tabular}
\end{table*}

\begin{figure*}
  \centering
  \includegraphics[height=6cm]{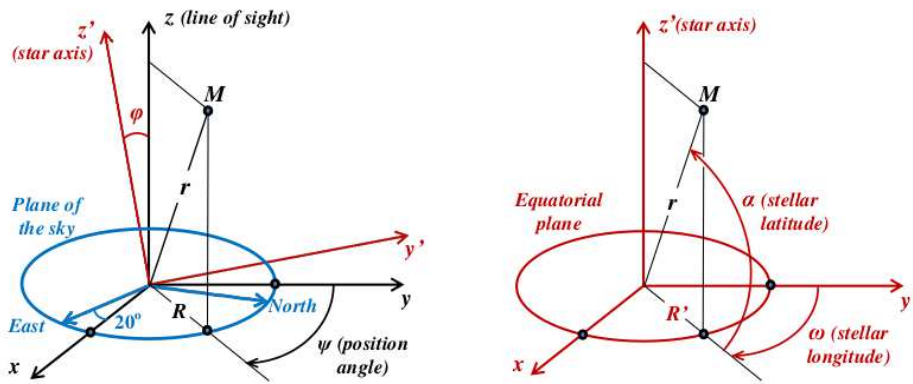}
  \caption{Left: sky coordinates; $x$ is in the plane of the sky, 20\dego\ north of east; $y$ is in the plane of the sky, 20\dego\ west of north; $z$ is along the line of sight; a vector \textbf{\textit{r}} pointing to M projects as $R$ on the sky plane, $R^2 =x^2 +y^2$ ; the position angle $\psi$ measured counter-clockwise from north, is such that $y=R\cos\psi$ and $x=R\sin\psi$. The stellar frame of coordinate $(x,y',z')$ is obtained from the sky frame $(x,y,z)$ by rotation of angle $\varphi$ about the $x$ axis. Right: stellar coordinates; a vector \textbf{\textit{r}} pointing to M projects on the equatorial plane $(x,y')$ as $R'=\sqrt{x^2 +y'^2}$; the stellar longitude $\omega$ is such that $y'=R'\cos\omega$ and $x=R'\sin\omega$; the stellar latitude $\alpha$ is such that $R'=r\cos\alpha$ and $z'=r\sin\alpha$.}
    
\label{fig1}
\end{figure*}

\begin{figure*}
\centering
\includegraphics[width=0.85\textwidth,trim=0.cm 0.5cm 0.cm 0.cm,clip]{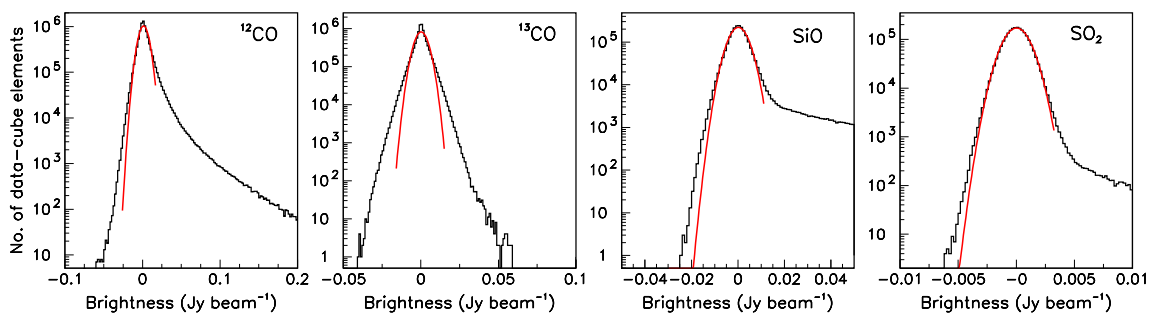}
\caption{Noise distributions within projected angular distances $R<12$, 10, 5, and 3 arcsec for $^{12}$CO(2-1), $^{13}$CO(2-1), SiO and SO$_2$ respectively (from left to right). The curves are Gaussian fits.}
\label{fig2}
\end{figure*}

The centre of the SO$_2$ line emission is offset by only $\sim$ 10 mas with respect to the centre of continuum emission and all observed line emissions share a common centre within 40 mas, consistent with measurement uncertainties; we therefore chose the origin of all maps at the centre of the SO$_2$ continuum map. 
As was done in \citet{Hoai2019} we use an orthonormal system of coordinates having the $y$ axis in the plane of the sky pointing 20\dego\ west of north and the $z$ axis pointing away from us along the line of sight. The $x$ axis points accordingly 20\dego north of east. This choice is such as to have the star axis defined in  \citet{Hoai2019} projecting on the sky plane along the $y$ axis. As will be seen below, such a choice is justified. Unless otherwise specified, we shall usually, for convenience, redefine north, south, east and west as halves of the sky plane having respectively $y>0$, $y<0$, $x>0$ and $x<0$. Figure \ref{fig1} summarizes the geometry and the coordinates being used. Figure \ref{fig2} displays the distributions of brightness for each of the observed lines, integrated over regions larger than the resolved source. Note the broad and non-Gaussian noise distribution for $^{13}$CO(2-1) observations.

\section{CO, S\lowercase{i}O and SO$_2$ emissions: comparing the main features} \label{sec3}
\begin{figure*}
\centering
\includegraphics[width=0.8\textwidth,trim=0.cm 0.5cm 0.cm 0.cm,clip]{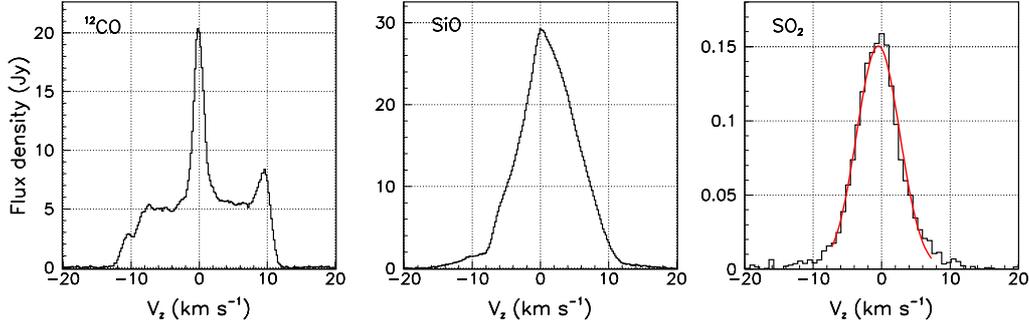}
\caption{From left to right: Doppler velocity distributions of CO, SiO and SO$_2$ emissions. CO and SiO spectra are integrated over $R<2.5$ arcsec and the SO$_2$ spectrum over $R<0.25$ arcsec. The line shown on the right panel is a Gaussian fit of 7.5 \kms\ FWHM.}
\label{fig3}
\end{figure*}

\begin{figure*}
\centering
\includegraphics[width=0.8\textwidth,trim=0.cm .5cm 0.cm 0.cm,clip]{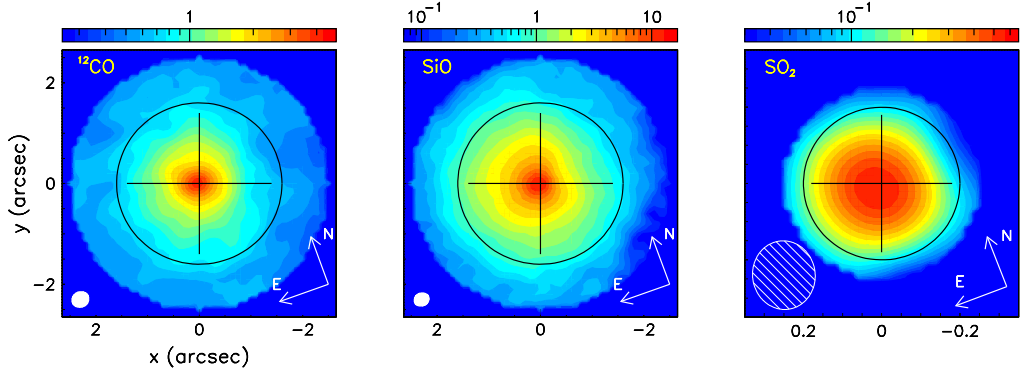}
\caption{From left to right: intensity maps of CO, SiO and SO$_2$ emissions. North is 20\dego\ east of the $y$ axis. Circles and crosses are meant to guide the eye. The beams are shown in the lower left corners of the panels. The colour scales are in units of Jy beam$^{-1}$\kms.}
\label{fig4}
\end{figure*}

\begin{figure*}
\centering
\includegraphics[width=0.9\textwidth,trim=0.cm 1.cm 0.cm 0.cm,clip]{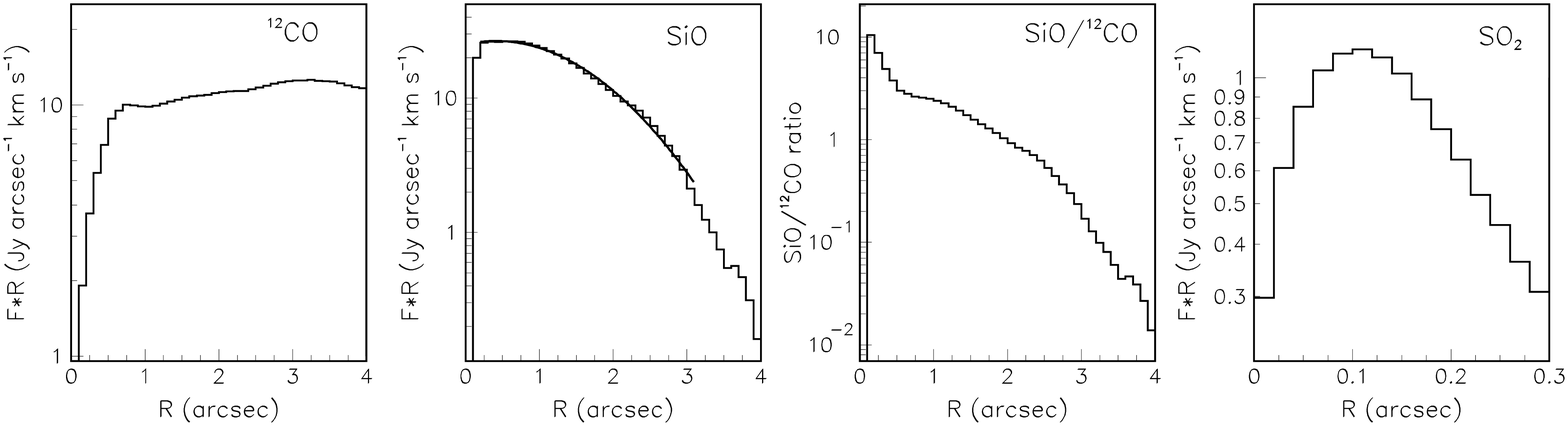}
\caption{Dependence on $R$ of the intensity multiplied by $R$ for $^{12}$CO, SiO, SiO/$^{12}$CO and SO$_2$ (from left to right). The SiO fit is a Gaussian centred at the origin and having a $\sigma$ of 1.2 arcsec.}
\label{fig5}
\end{figure*}

CO, SiO and SO$_2$ emissions probe very different ranges of $r$, the distance to the central star. While CO molecules are formed early and survive up to long distances, being only affected by UV dissociation from the interstellar radiation field \citep{Mamon1988, Groenewegen2017}, SiO molecules tend to aggregate in grains, causing their progressive disappearance from the gas \citep{Gonzalez2003, VandeSande2018} at distances at arcsec scale from the star in addition to their early dissociation from interstellar UV radiation \citep[]{Willacy1997}; and while both CO and SiO emissions are excited by collisions with other molecules, in majority hydrogen, SO$_2$ emission is mostly excited by absorbing the UV radiation of the central star, confining it to its very close neighbourhood \citep{Yamamura1999, Danilovich2016}. As the CO emission of the circumstellar envelope at distances exceeding $\sim$ 250 au was studied in detail in \citet{Hoai2019} and as SiO and SO$_2$ emissions are confined to smaller distances, we restrict the present analysis to angular distances (projected on the sky plane) $R<2.5$ arcsec. In the case of SO$_2$ emission, a more restrictive cut at $R=0.25$ arcsec is normally applied in order to stay above noise. Doppler velocity spectra and intensity maps are illustrated in Figures \ref{fig3} and \ref{fig4} respectively. Figure \ref{fig5} displays the dependence on $R$ of the measured intensities and compares that of SiO with that of CO used as reference. 

The diversity of the Doppler velocity spectra displayed in Figure \ref{fig3} results from that of the radial ranges that the associated emissions are exploring. Very close to the star, we shall argue in the next section that the SO$_2$ line profile may receive significant contribution from processes other than the dominant coherent Doppler broadening associated with the large beam size, the wind being too slow to produce the broad spectrum that is observed. In \citet{Hoai2019}  we remarked that thermal broadening, $\sqrt{2kT/M_{CO}}$ where $T$ is the temperature, $k$ Boltzmann constant and $M_{CO}$ the mass of the CO molecule, amounts to only 0.16 \kms\ at $T$=45 K. As $M_{SO_2}\sim2.3 M_{CO}$ and as temperature is now expected to reach values at the 400 to 600 K level, we expect a thermal broadening of $\sim$ 0.3 to 0.4 \kms, still considerably smaller than the observed width. A similar comment applies to the SiO profile, which, however, receives important contribution from coherent Doppler broadening of the nascent wind.

The intensity maps shown in Figure \ref{fig4} display approximate isotropy. The influence of temperature on the $r$-dependence of the effective emissivity is governed by a factor $Q=Q_0(2J+1)(2.8/T)\exp(-E_J/T)$ where the values of $J$ and $E_J$ are listed in Table \ref{tab2} for each of the four lines together with the Einstein coefficients to which $Q_0$ is proportional. Neglecting absorption and assuming the $r$-dependence of the temperature obtained by \citet{Hoai2019} from a comparison between $^{12}$CO(1-0) and $^{12}$CO(2-1) emissions, the CO and SiO effective emissivity has decreased by only 13\% and respectively 22\% when $r$ reaches 2 arcsec. The important observed decline of SiO emission with respect to CO emission, by a factor of nearly 20 when $R$ reaches 2.5 arcsec (middle right panel of Figure \ref{fig5}), is therefore evidence for a decrease of the SiO emission other than expected from the temperature gradient, probably due to a decrease of the gas density resulting from a combination of UV dissociation and accretion on grains. Moreover, the deviation of the SiO/CO ratio from a simple exponential, with a sharper decrease in the first half of an arcsec, also requires an explanation other than being caused by a temperature gradient. We shall see later that it is associated with the presence of narrow polar jets (meaning collimated streams having velocity significantly larger than the surrounding gas)  but a quantitative evaluation is premature at this stage: Figure \ref{fig5} illustrates the dependence of the intensity on $R$ rather than on $r$, including therefore contributions from large $r$ values that would require de-projection to be taken in proper account.

\begin{table*}
  \centering
  \caption{Temperature dependence of the effective emissivity.}
  \label{tab2}
  \begin{tabular}{cccccc}
    \hline
    \multirow{2}{*}{Molecule} &\multicolumn{2}{c}{Quantum numbers}    &\multirow{2}{*}{Einstein coefficients (Hz)$^*$} &\multirow{2}{*}{Rest frequency (GHz)}& \multirow{2}{*}{$E_J$ (K)} \\
    
   &Vibrational
    &Rotational&&&\\
    \hline
    $^{13}$CO    &$\nu$=0     &$J$=2-1     &6.038 10$^{-7}$      &220.399     &15.9\\    
    $^{12}$CO     &$\nu$=0     &$J$=2-1     &6.910 10$^{-7}$     &230.538     &16.6\\
    SiO     &$\nu$=0     &$J$=5-4     &5.917 10$^{-4}$     &217.105     &31.26\\
    SO$_2$     &$\nu$=0     &16$_{6,10}$-17$_{5,13}$ $^{**}$     &2.349 10$^{-5}$     &234.422      &213.32\\
    \hline

\multicolumn{6}{l}{$^*$ From \citet{Schoier2005}} \\
\multicolumn{6}{l}{$^{**}$ This is erroneously quoted by \citet{Homan2018} as 28$_{3,25}$-28$_{2,26}$} \\
  \end{tabular} \\
\end{table*}

Not clearly apparent on Figures \ref{fig3} and \ref{fig4} is the strong asymmetry of the SiO and 
 SO$_2$ data cubes, which is revealed by comparing the flux densities integrated in each of the octants obtained by splitting red-shifted/blue-shifted, north/south and east/west (defined in rotated coordinates). Table \ref{tab4} summarizes these results by averaging over north and south. A toy model parameterization in terms of two parameters $u$ and $v$, respectively measuring the blue/red and east/west asymmetries, gives a good fit to the CO and SiO data but not to the SO$_2$ data. It uses two parameters to describe four quantities related by one relation (their sum is equal to 4): blue-east=$1+u+v$, blue-west=$1+u-v$, red-east=$1-u+v$ and red-west=$1-u-v$. The values of $u$ and $v$ are listed in the table together with the rms deviation between model and observed values.

\begin{table*}
  \centering
  \caption{Normalised brightness of the $^{12}$CO(2-1), SiO and SO$_2$ data-cube quadrants. East and west mean $x>0$ and $x<0$ respectively.}
  \label{tab4}
  \begin{tabular}{ccccccc}
\hline
&\multicolumn{2}{c}{CO}
&\multicolumn{2}{c}{SiO}
&\multicolumn{2}{c}{SO$_2$} \\

&East
&West
&East
&West
&East
&West \\
    \hline
Blue
&1.02
&0.86
&0.86
&0.46
&1.48
&0.82\\
Red
&1.09
&1.04
&1.55
&1.12
&0.81
&0.90\\
$u$
&\multicolumn{2}{c}{$-$0.06}
&\multicolumn{2}{c}{$-$0.34}
&\multicolumn{2}{c}{0.15}\\
$v$
&\multicolumn{2}{c}{0.05}
&\multicolumn{2}{c}{0.21}
&\multicolumn{2}{c}{0.14}\\
Rms
&\multicolumn{2}{c}{0.03}
&\multicolumn{2}{c}{0.01}
&\multicolumn{2}{c}{0.19}\\
\hline
  \end{tabular}
  \end{table*}

The very large asymmetry displayed by the SiO data-cube is characterized by a strong depression of the blue-western quadrant with respect to its red-eastern counterpart. It is interesting to note that the CO asymmetry, although much smaller, is of the same nature, suggesting that it may have the same source as the SiO asymmetry, its effect being much stronger at short distances from the star. At variance with CO and SiO emissions, SO$_2$ emission is dominated by a blue-red asymmetry in the eastern hemisphere and is not properly described by the toy model parameterization, revealing the very different morpho-kinematic regime probed by this line close to the star.

We remark that both expansion and rotation, or for that matter any linear combination of those, should produce a centrally symmetric data-cube. The strong central asymmetry of the SiO data-cube is therefore evidence for something else than central expansion or rotation. The centrally symmetric component of the data-cube, $f_S(x,y,V_z)=\frac{1}{2}[f(x,y,V_z)+f(-x,-y,-V_z)]$ should keep track of the respective roles of rotation and expansion in the morpho-kinematics of the circumstellar envelope, leaving aside whatever is causing the strong central asymmetry; in the centrally symmetric data-cube, the rms deviation of the octant flux densities with respect to their mean is only 5\% instead of 40\% for the original data-cube (note that the value of 5\% gives an upper limit to the uncertainties attached to the normalised brightness values listed in Table \ref{tab4}).

\section{SO$_2$ emission} \label{sec4}

The confinement of SO$_2$ emission to the immediate neighbourhood of the star, less than 30 au, makes it a very precious source of information about the mass loss mechanism in its earlier phase. SO$_2$ is known to trace hot gas in the immediate neighbourhood of oxygen-rich AGB stars \citep{Yamamura1999} and its emission is mostly radiatively excited \citep{Danilovich2016}. The temperature has been measured by \citet{Hoai2019} from a comparison of $^{12}$CO(1-0) and $^{12}$CO(2-1) emissions at distances from the star exceeding $\sim$ 2 arcsec; they proposed two different forms for the radial dependence: $T\sim109 \exp(-r/3.1)$ and $T\sim106/r$ (with $T$ in Kelvin and $r$ in arcsec). While the former gives the best fit at large values of $r$, the latter is better adapted to extrapolation toward small values of $r$, suggesting that the gas temperature in the region probed by SO$_2$ is reaching a few hundred Kelvin, in agreement with the estimate of \citet{Yamamura1999} for other oxygen-rich AGB stars. While the details of the formation, emission and destruction of SO$_2$ molecules in the neighbourhood of oxygen-rich AGB stars are not fully understood, a number of features have been clearly established \citep{Yamamura1999, Cherchneff2006, Danilovich2016, Gobrecht2016}: they trace warm ($\sim$600 K) gas layers hosting turbulence and shocks produced by the star pulsations that favour their formation through the liberation of atomic oxygen; they absorb UV photons from the star that cause both their excitation and dissociation with the result that emission is confined to a narrow radial range, typically Gaussian with $\sigma$ values in the range of a few tens of au; line widths (FWHM) are typically at the level of 5 to 10 \kms, as also observed in star forming regions at lower temperatures \citep{Esplugues2013}. It is therefore natural to expect that the line profile in the present observations receives important contribution from such effects, competing with coherent Doppler broadening caused by rotation and by the expansion of the nascent wind. While the beam size (0.18$\times$0.17 arcsec$^2$) and the broad line width smear the data-cube in a way that prevents a detailed study of the morpho-kinematics, a number of conclusions can be drawn from closer inspection of the data-cube. 

\begin{figure*}
\centering
\includegraphics[height=4.5cm,trim=0.cm .5cm 0.cm 0.cm,clip]{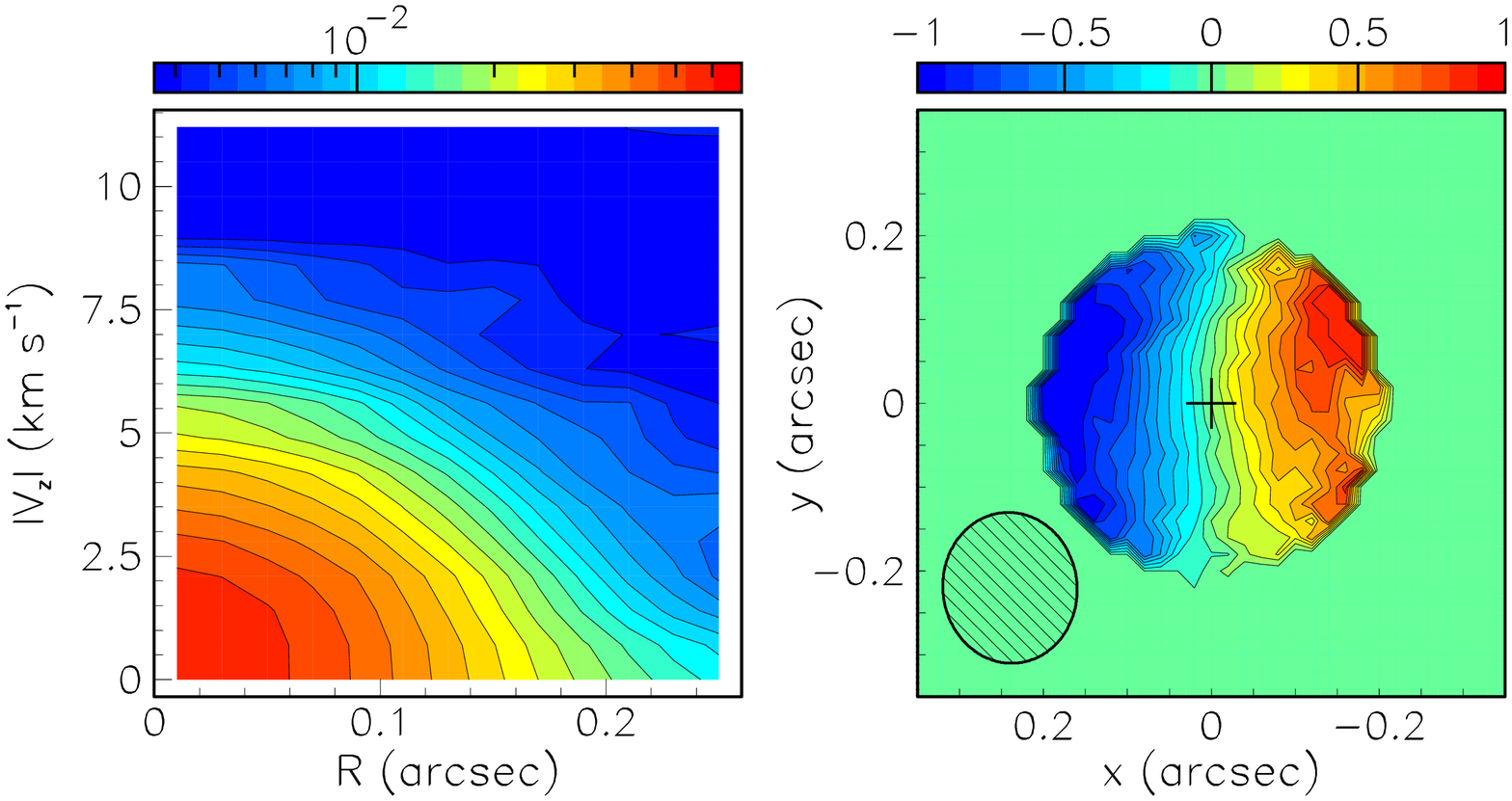}
\includegraphics[height=4.5cm,trim=0cm .5cm 1cm 0.cm,clip]{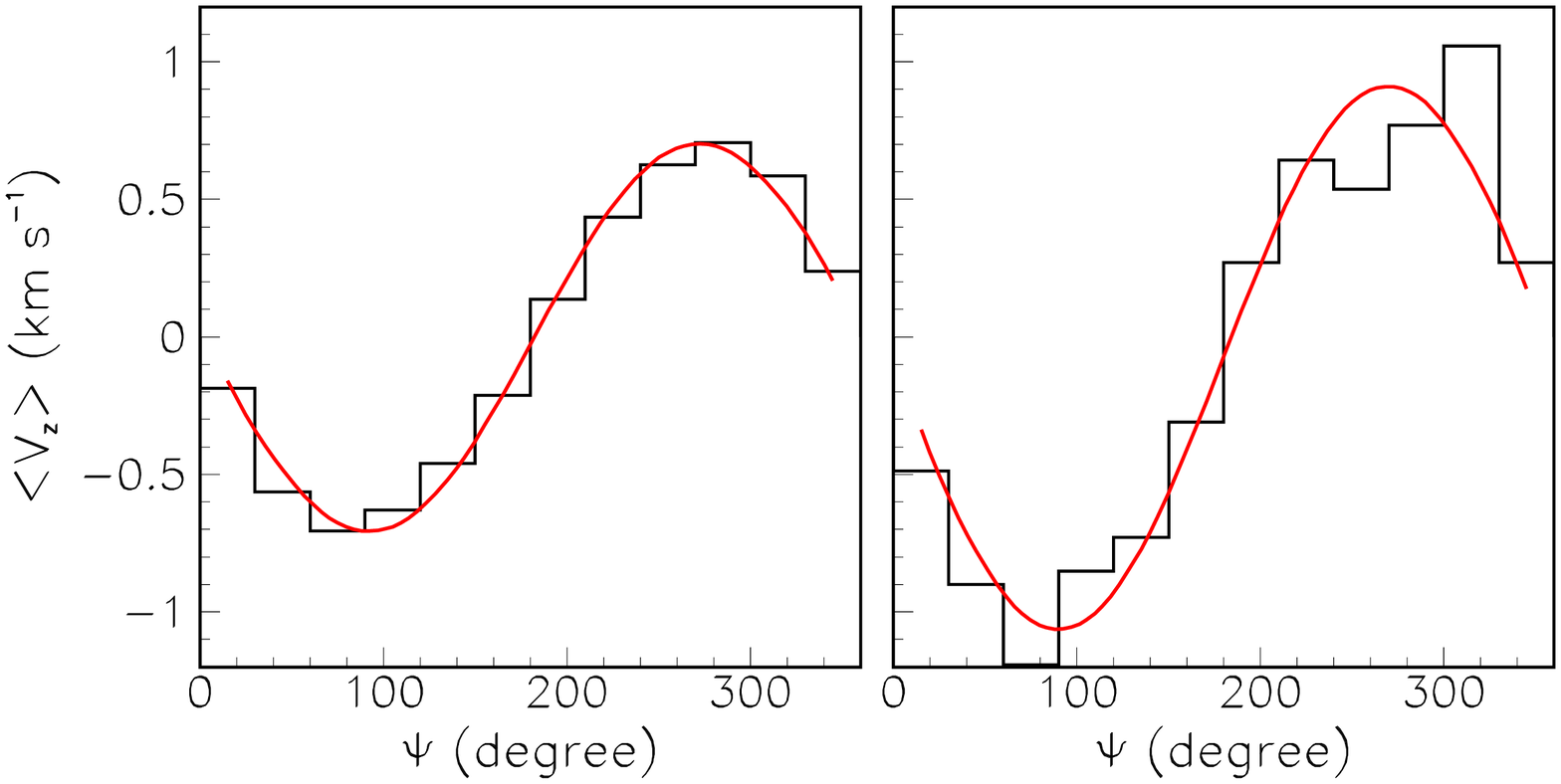}
\caption{SO$_2$ emission (3-$\sigma$ noise cut). Left: P-V map of $|V_z|$ vs $R$. The colour scale is in units of Jy beam$^{-1}$. Middle left: sky map of the mean Doppler velocity (\kms, $R<0.2$ arcsec). Right panels: dependence of the Doppler velocity on position angle $\psi$ for $0.05<R<0.15$ arcsec (middle right) and $0.15<R<0.25$ arcsec (right).}
\label{fig6}
\end{figure*}

Figure \ref{fig6} (left panel) displays the position-velocity (P-V) map of $|V_z|$ as a function of $R$. It reveals no particularly remarkable feature, such as a polar or equatorial enhancement, and how much radial expansion it implies depends on the amount of smearing caused by the broad line width. To obtain some rough evaluation of the line width, we calculate the width (FWHM) of the line profile observed in each pixel as rms deviation with respect to the mean. We find that it decreases from $\sim$ 7.7 \kms\ for $R<0.15$ arcsec to $\sim$ 5.6 \kms\ for $R>0.15$ arcsec. To separate different contributions to the line width would require a significantly better resolution; we simply retain, as an order of magnitude, that both radial expansion velocity and other possible sources of line broadening are at the scale of 5 \kms\ and cannot significantly exceed 7 \kms.

The map of the mean Doppler velocity, displayed in Figure \ref{fig6} (middle left panel), suggests the presence of rotation, as previously remarked by \citet{Homan2018}. 

Figure \ref{fig6} (right panels) displays distributions of the Doppler velocity as a function of position angle $\psi$ measured counter-clockwise from the $y$ axis. In spite of the large beam size a significant comparison can be made between two radial intervals, below and above 0.15 arcsec. Sine wave fits give respective amplitudes of 0.70 and 0.99 \kms\ and phase shifts that nearly cancel. The amplitude obtained for the whole $R$ range is 0.81 \kms.  The small phase shifts, meaning that the axis of rotation projects on the sky plane along the $y$ axis, imply also the absence of a significant anisotropy, polar or equatorial, of the expansion velocity \citep{Diep2016}. We retain from this a mean rotation velocity of approximately $0.81/\sin10^\circ \sim 4$ to 5 \kms\ . This number is but a rough evaluation of the scale of the velocities that are at stake but provides evidence for a significant rotation component of the velocity field.

Additional support is provided by the observation that the separation in $x$ between the blue-shifted and red-shifted components is significantly non-zero and persists up to the larger values of $|V_z|$.

In summary, the observation of SO$_2$ emission, when limited to a projected distance from the star $R<\sim0.25$ arcsec, is consistent with a combination of rotation and isotropic radial expansion confined to less than $\sim$ 30 au from the star; rotation velocities are of the order of 4 to 5 \kms\ and radial expansion velocities increase to a few \kms\ with no evidence for departure from isotropy; both contribute to the observed line width of $\sim$ 7.5 \kms\ FWHM, which may also receive significant contribution from turbulent Doppler broadening.

\section{S\lowercase{i}O emission} \label{sec5}

\subsection{General remarks} \label{sec5.1}

\begin{figure*}
\centering
\includegraphics[width=0.8\textwidth,trim=0.cm 0.5cm 0.cm 0.cm,clip]{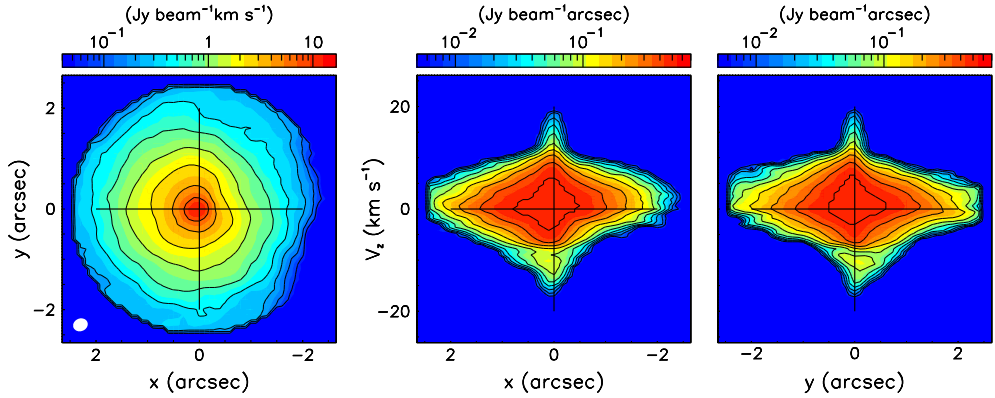}
\caption{Projections of the SiO data-cube on ($x,y$) (left), ($x,V_z$) (middle) and ($y,V_z$) (right).}
\label{fig9}
\end{figure*}

\begin{figure*}
\centering
\includegraphics[width=0.8\textwidth,trim=0.cm 0.5cm 0.cm 0.cm,clip]{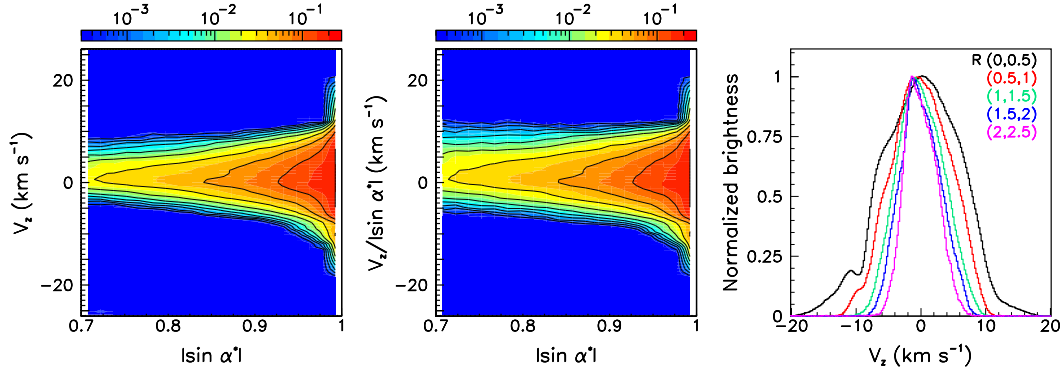}
\caption{SiO emission. Left: Projection of the SiO data-cube on the ($|\sin\alpha^*|,V_z$) plane with $|\sin\alpha^*|=\sqrt{r^{*2}-R^2}/r^*$ and $r^*=3.5$ arcsec. The boundary displays the dependence of $V_z$ on $\alpha$ at $r=r^*$.  Middle: same as left with $V_z$ replaced by $V=V_z/|\sin\alpha^*|$. The colour scales are in units of Jy beam$^{-1}$. Right: re-centred line profiles normalized to a same peak value for different intervals of $R$ (see insert).}
\label{fig10}
\end{figure*}

SiO emission probes the radial range between the immediate neighbourhood of the star, where rotation is present, and the outer part of the circumstellar envelope, dominated by expansion. As was shown in Figure \ref{fig5}, a remarkable feature of SiO emission is its short radial range, $\sim 3$ arcsec. This has important consequences on the structure of the data-cube: its projections on the $(x,V_z)$ and $(y,V_z)$ planes, which are P-V maps, display sharp boundaries associated with this short radial range. They are shown in Figure \ref{fig9} together with the projection of the data-cube on the $(x,y)$ plane (intensity map). Neglecting the inclination $\varphi$ of the star axis with respect to the line of sight, the ratio $R/r$ measures the cosine of the stellar latitude $\alpha$ and the position velocity maps can be redrawn in the $(R,V_z)$ plane, their boundary displaying the maximal value of $V_z$ as a function of $R$. Defining $\alpha^*=\cos^{-1}(R/r^*)$ with $r^*$ being the radial range of SiO emission, the projection of the data-cube on the $(\sin\alpha^*,V_z)$ plane displays therefore the dependence of $V_z$ on $\alpha$ at $r=r^*$. This is shown in Figure \ref{fig10}, using a 3-$\sigma$ noise cut on $f$ and assuming that $r^*$ is the same in all directions (and taken equal to 3.5 arcsec); also shown is the projection of the data-cube on the $(\sin\alpha^*,V_z/\sin\alpha^*)$ plane, displaying instead the dependence on $\alpha$ of the radial expansion velocity $V=V_z/\sin\alpha^*$ at $r=r^*$. Note that the deviation of $\varphi$ from zero, of the order of $\sim$ 10\dego, simply smears the boundaries of the P-V maps without significantly affecting the result. Under such assumptions, the P-V maps of Figure \ref{fig10} cover latitudes between $\sim 45^\circ [\cos^{-1}(2.5/3.5)]$ and 90\dego\ and provide direct evidence for three important features: 1) two narrow polar jets with velocity reaching $\sim$ 20 \kms, covering an interval of $\sim$ 0.015 units of $\sin\alpha^*$ below 1, meaning an opening angle of $\sim \pm10^\circ$; additional material related to the jet properties is presented in Appendix \ref{appa1}; 2) a slower wind with maximal radial velocity of $\sim$ 12 \kms\ decreasing very slowly with latitude in the red hemisphere, its counterpart in the blue hemisphere having a maximal velocity decreasing faster with latitude down to $\sim$ 5 \kms; 3) an important asymmetry between the blue and red hemispheres, consistent with the observations previously made in Section \ref{sec3}.

However, the above arguments assume that the radial range $r^*$ of SiO emission is the same in all directions. But whatever is causing the wind to speed up and whatever is causing the emissivity to decrease must be related, if not directly at least as being both dependent on the physical parameters defining the state of the environment (density, temperature, turbulence and shocks, etc...). The strong red-east/blue-west asymmetry evidenced in the middle panel of Figure \ref{fig9} shows that the wind velocity is low at the edge of the emissivity region in the blue-western quadrant; but this may be because acceleration is less efficient in the blue-western quadrant or because the edge of emissivity is closer to the star in the blue-western quadrant. This ambiguity is inherent to the nature of the observations and must be kept in mind when seeking an interpretation. Yet, the P-V maps of Figures \ref{fig9} and \ref{fig10} show that the terminal velocity of $\sim$ 12 \kms\ observed in CO emission at large distances from the star \citep{Hoai2019} has already been reached in the red hemisphere within the range explored by SiO emission; they also provide evidence for the wind being accelerated at distances from the star ranging between $\sim$ 50 au and $\sim$ 300 au. To say more at this stage, without a preliminary understanding of the mechanism of acceleration and of grain formation, is not possible: we postpone further comments to Section \ref{sec7}. Several scenarios have been proposed to launch the nascent wind: direct acceleration from star pulsations \citep{McDonald2016}, magnetic fields \citep{Vlemmings2013}, stellar rotation \citep{GarciaSegura1999, GarciaSegura2014}, photon collisions on transparent silicate grains \citep{Woitke2006, Hofner2008} or absorption of the star UV light by standard dust grains \citep[for a recent review see][]{Hofner2018}; none of these, on its own, can generate an asymmetry of the type observed here. 
 
As was done for SO$_2$ emission, we obtain some rough evaluation of the line width, more precisely of an upper limit to the line width, by calculating the width of the line profile observed in each pixel as rms deviation with respect to the mean. The result is displayed in the right panel of Figure \ref{fig10} and gives evidence for a clear decrease of the line width (FWHM) as a function of $R$: respectively 13.6, 10.2, 8.1, 7.0 and 5.8 \kms\ for 0.5 arcsec wide intervals covering from 0 to 2.5 arcsec. While giving evidence for a major contribution of coherent Doppler broadening, this result leaves room for a significant contribution of turbulence.

\subsection{Detailed description of the asymmetry of the data-cube} \label{sec5.2}
\begin{figure*}
\centering
\includegraphics[width=0.9\textwidth,trim=0.cm 1.cm 0.cm 0.cm,clip]{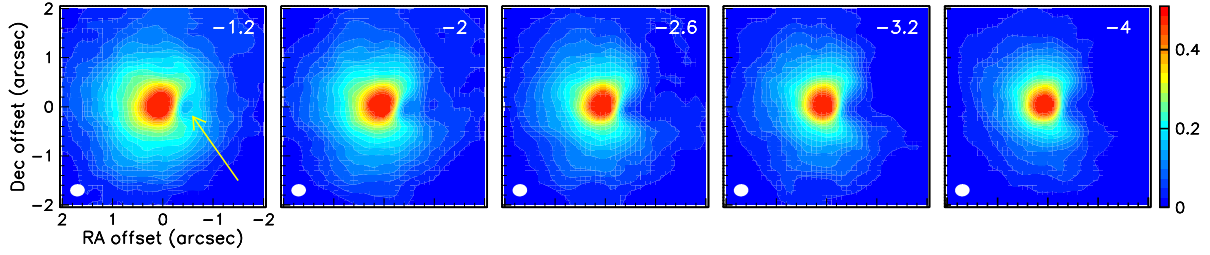}
\caption{SiO emission. Channel maps at $V_z=-1.2$, $-$2, $-$2.6, $-$3.2 and $-$4 \kms\ as indicated in the inserts. The arrow in the left panel points to the feature associated by \citet{Homan2018} with a possible companion. To ease comparison with \citet{Homan2018} the maps are in sky coordinates, with north pointing up. The colour scale is in units of Jy beam$^{-1}$.}
\label{fig11}
\end{figure*}

The evidence for narrow polar jets brings with it two major questions: which is the jet-launching mechanism? and which role do the jets play in the generation of the wind? The morpho-kinematics of SiO emission has been interpreted by \citet{Homan2018} in terms of the presence of a companion gravitationally attracting gas around it. These authors made a sign mistake when comparing Doppler velocities between SiO and CO emissions, invalidating their assertion that a lane of gas is bridging the gap between the star and its hypothetical companion (right panel of their Figure 13). Yet, their observation of a nearly point-like void in the channel maps of SiO emission, at $\sim$ 0.5 arcsec west of the star and covering a broad range of Doppler velocities in the red hemisphere (meaning in fact the blue hemisphere because of the sign mistake) remains valid. This is illustrated in Figure \ref{fig11} that displays channel maps in the relevant range of $V_z$ \citep[contrary to the rest of the article, we use here sky coordinates with north pointing up in order to ease the comparison with][]{Homan2018}. In spite of the slightly larger beam size \citep[0.29$\times$0.25 arcsec$^2$ instead of 0.20$\times$0.18 arcsec$^2$ for][]{Homan2018} the void is clearly visible and is seen to leave room for a significant depletion when moving further out in the blue-shifted direction.

\begin{figure*}
\centering
\includegraphics[width=0.8\textwidth,trim=0.cm 0.5cm 0.cm 0.cm,clip]{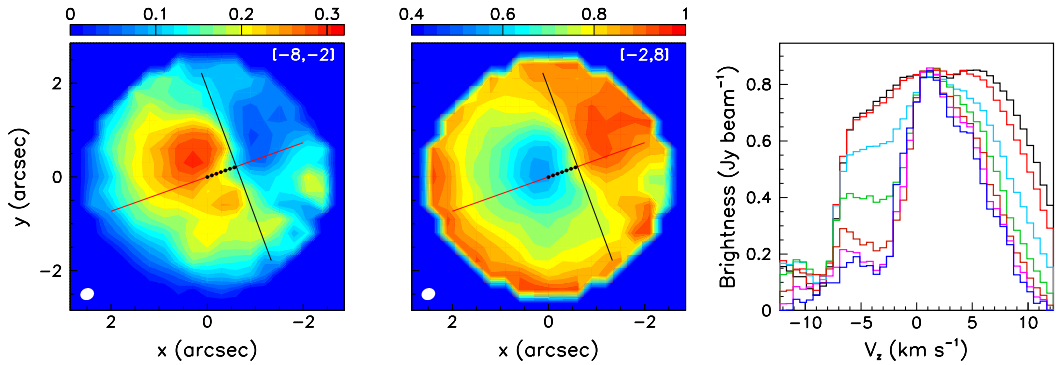}
\caption{SiO emission. Channel maps in $-8<V_z<-2$ \kms\ (left) and $-2<V_z<8$ \kms\ (middle) intervals normalized to the whole $V_z$ range. Black lines delineate the sharp transition to the depleted region. Right: individual pixel Doppler velocity spectra obtained by scanning in steps of 0.1 arcsec along the red line depicted in the left panels from ($x,y$)=(0,0) arcsec (black) to ($x,y$)=(0.6,0) arcsec (blue) in sky coordinates (not rotated by 20\dego). }
\label{fig12}
\end{figure*}

In order to better understand the nature of the above feature, we display in Figure \ref{fig12} (this time turning back to rotated coordinates with north pointing 20\dego\ east of the $y$ axis) channel maps associated with two much broader intervals of $V_z$: respectively $-8<V_z<-2$ \kms\ and $-2<V_z<8$ \kms. Moreover we normalize the maps by dividing the intensity measured in the interval of $V_z$ by its value over the whole $V_z$ range. This reveals the sharp transition between data-cube elements located in the eastern and western hemispheres. Scanning toward the boundary displays a spectacular evolution of the Doppler velocity distribution progressively emptying the blue-shifted hemisphere when moving westward.

\begin{figure*}
\centering
\includegraphics[width=0.49\textwidth,trim=0.cm .5cm 0.cm 2.cm,clip]{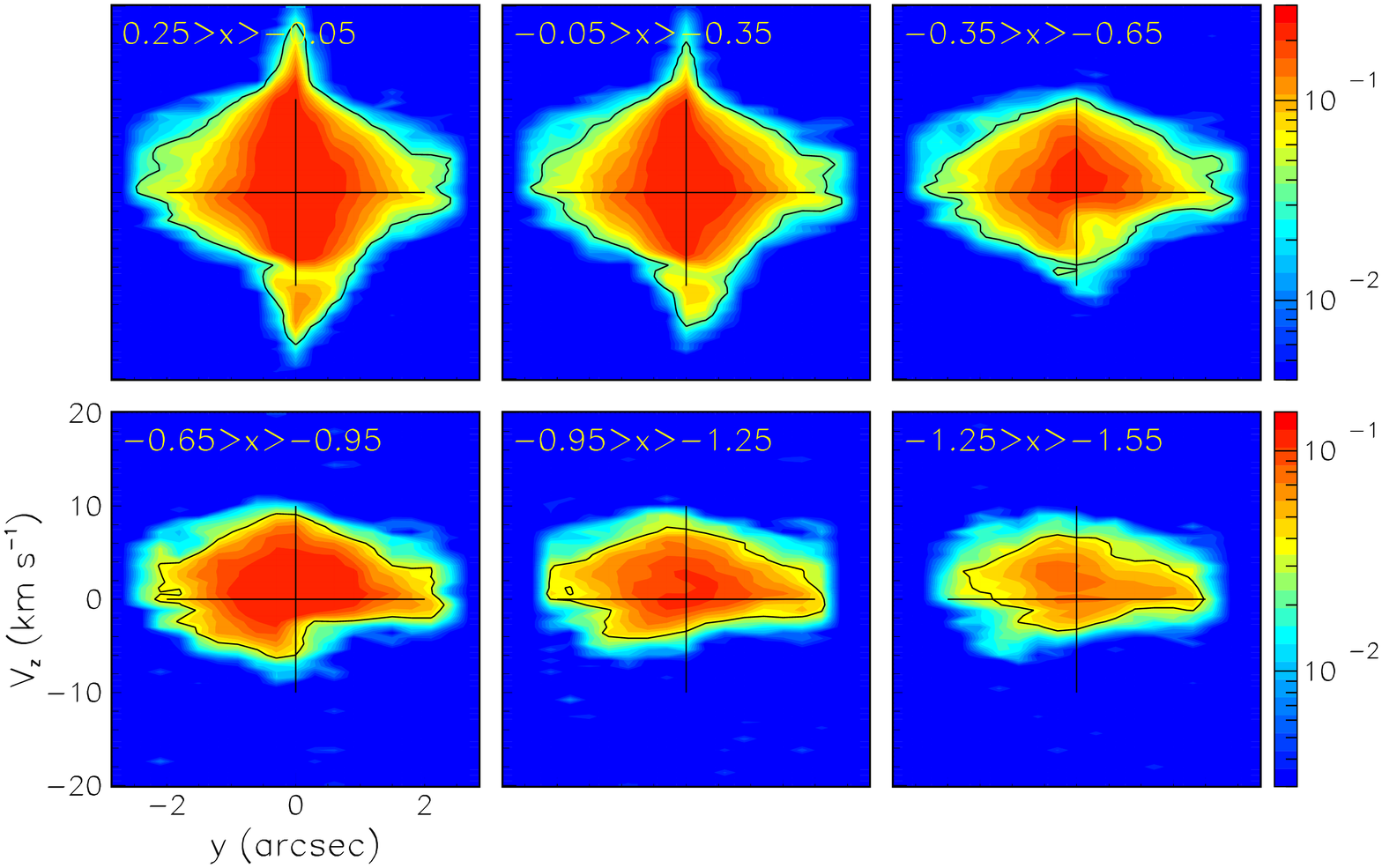}
\includegraphics[width=0.49\textwidth,trim=0.cm 0.5cm 0.cm 2.cm,clip]{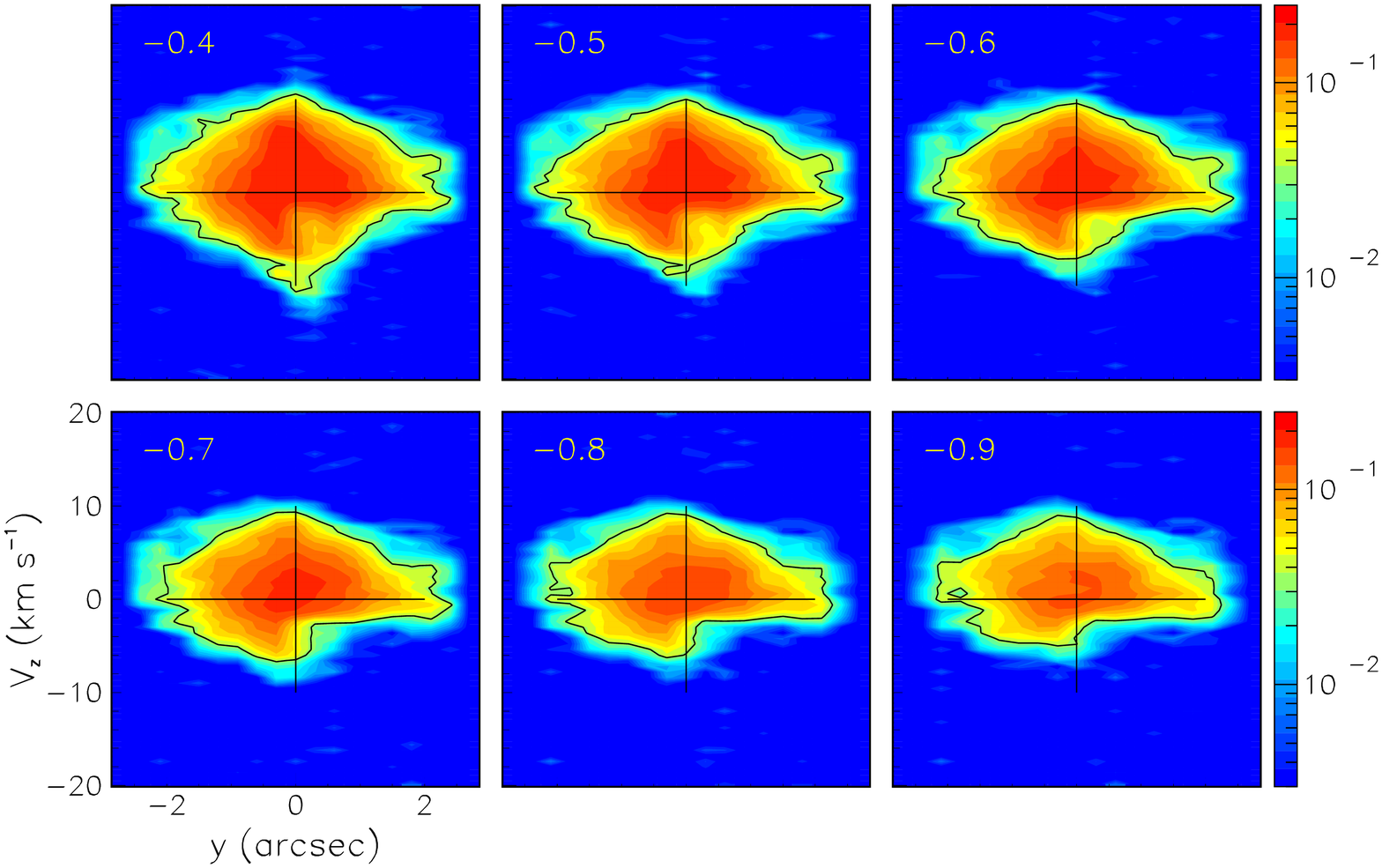}
\caption{SiO emission. P-V maps in various intervals of $x$ indicated in the inserts. The left panels are for 0.3 arcsec wide $x$ intervals covering from 0.3 to $-$1.5 arcsec. The right panels are for 0.1 arcsec wide intervals (one pixel) centred at $-$0.4 to $-$0.9 arcsec. The colour scales are in units of Jy beam$^{-1}$.}
\label{fig13}
\end{figure*}

An overall picture of the asymmetry evidenced in Figures \ref{fig11} and \ref{fig12} is presented in Figure \ref{fig13}, which scans across the data-cube in narrow slices of $x$ in the region where the asymmetry has been revealed. When scanning westward, approximate symmetry in the $(V_z,y)$ plane is observed up to $x \sim -0.3$ arcsec where a depression appears at $(V_z, y)\sim(-4$ \kms, 0.3 arcsec). In an $x$ interval of only 0.4 arcsec, this depression expands very rapidly to the whole south-western-blue octant: by $x=-0.7$ arcsec the region $x<-0.7$ arcsec, $y>0$, $V_z<-2$ \kms\ has been completely depleted. It then slowly expands into the south-eastern-blue octant. This depression dominates the asymmetry and is outstanding both because of its large amplitude and because of the sharpness of its boundary. The similarity between its morphology and that observed near QX Pup by \citet{SanchezContreras2018} suggests that we are looking at a similar phenomenon. Their interpretation is that the bow-shock of a bipolar SiO outflow launched at some 10 to 20 \kms\ (probably by a very close-by invisible companion) is carving a cavity into the ambient gas. A similar observation is reported in \citet{Sahai2006} who study the molecular flow of the pre-planetary nebula IRAS 22036+5306, however this time with a much faster wind. Additional material related to the asymmetry of the data-cube is presented in Appendix \ref{appa2}. 

\subsection{Rotation}\label{sec5.3}

In Section \ref{sec4} evidence was given for rotation at small distances from the star; as expansion dominates at large distances, it is therefore natural to expect that the wind velocity evolves to nearly pure expansion in the range of distances explored by SiO emission.

Both rotation and expansion produce centrally symmetric data-cubes, the former producing an east-west asymmetry of the mean Doppler velocity, the latter a north-south asymmetry. A major difference between the two is that rotation produces opposite redshifts on opposite sides of the axis; in the present case the whole eastern hemisphere is blue-shifted and the whole western hemisphere is red-shifted as seen in the case of SO$_2$ (Figure \ref{fig6}). In the case of expansion, equator and poles produce Doppler shifts of opposite signs in a same hemisphere, north or south: the map of the mean Doppler velocity, $\langle V_z\rangle$, depends on how prolate, or oblate, is the velocity latitudinal distribution. In particular, a spherical distribution trivially causes $\langle V_z\rangle$ to cancel everywhere on the sky map. This feature was used in \citet{Hoai2019} to give evidence for dominance of expansion when comparing the broad and narrow components of CO emission at large distances from the star. A consequence is that rotation is more efficient than expansion at generating an asymmetry of the $\langle V_z\rangle$ map and, in the present case, the strong blue-west depletion prevents a reliable distinction between expansion and rotation by introducing a bias that systematically favours an interpretation in terms of rotation. This is illustrated in Section \ref{appa3} of the appendix by using a centrally symmetrized data-cube. In the present section we simply limit the analysis to the red-shifted hemisphere, assuming that it is not significantly affected by the depletion.

\begin{table*}
  \centering
  \caption{Fits of the form $\langle V_z \rangle=V_{z0}+V_{z1}\cos(\psi+\psi_0)$ to the dependences on position angle displayed in Figure \ref{fig14}.}
  \label{tab6}
  \begin{tabular}{ccccc}
    \hline
    $V_z$ range (\kms)&$R$ range (arcsec)& $V_{z0}$ (\kms) & $V_{z1}$ (\kms) & $\psi_0$ (deg)\\
    \hline
    \multirow{2}{*}{2 to 5} & 0.5 to 1.5 & 3.40&$-$0.03&$-$3\\
     \smallskip
    & 1.5 to 2.5 &3.31&$-$0.06&12\\
    \multirow{2}{*}{5 to 10}&0.5 to 1.5&6.65&$-$0.31&25\\
     \smallskip
     &1.5 to 2.5 &6.08&$-$0.35&17\\
     \multirow{2}{*}{0 to 20}&0.5 to 1.5&3.35&$-$0.36&11\\
     \smallskip
     &1.5 to 2.5&2.80&$-$0.31&$-$10\\
     $>0^*$&$<8^*$&4.7$^*$&$-$0.34$^*$&$-$4$^*$\\
      \hline
      \multicolumn{5}{c}{$^*$ Red-shifted CO broad component, \citet{Hoai2019}}
  \end{tabular}

\end{table*}

\begin{figure*}
\centering
\includegraphics[width=0.6\textwidth,trim=0.cm 0.5cm 0.cm 0.cm,clip]{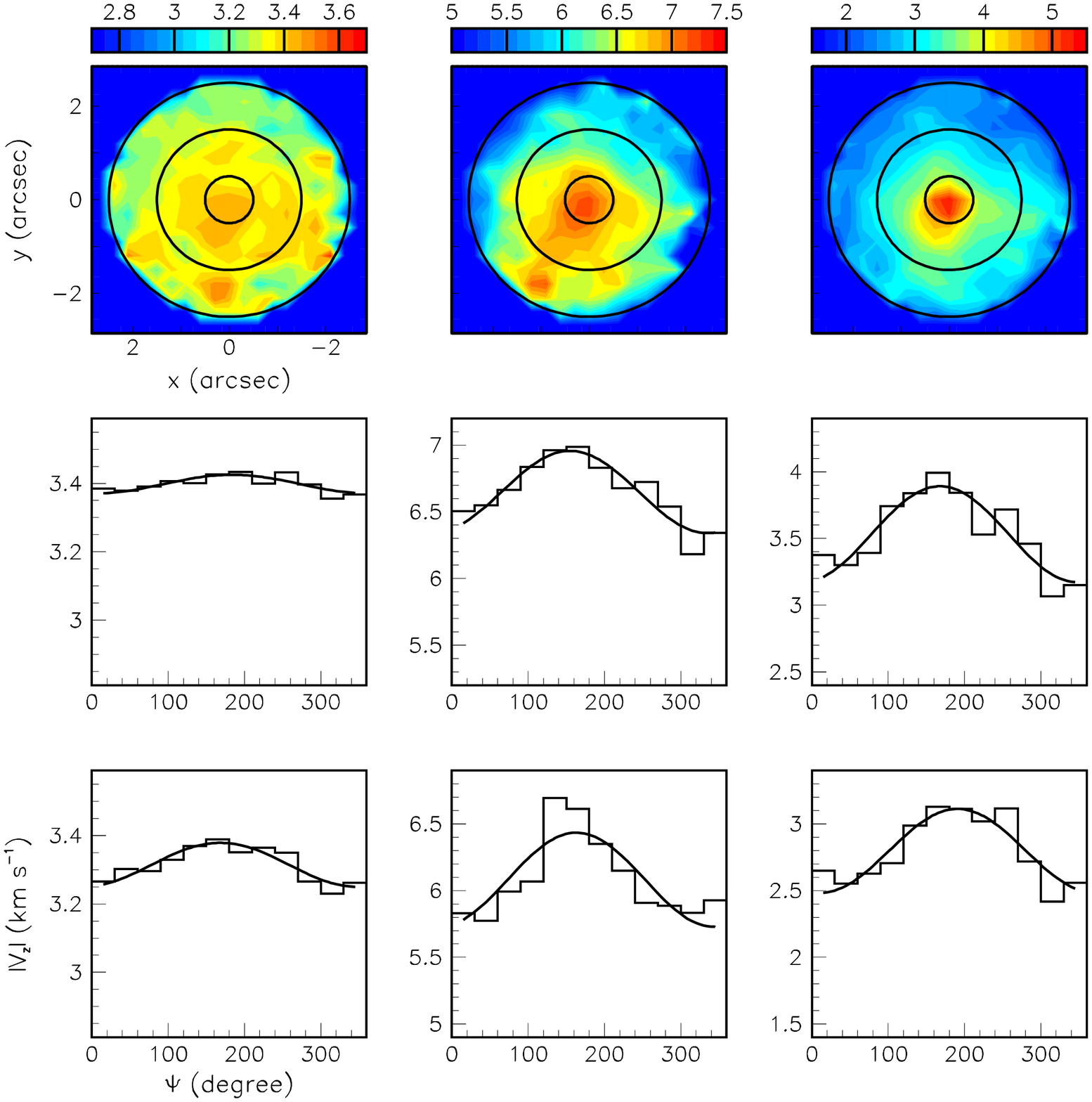}
\caption{The mean Doppler velocity $\langle V_z \rangle$ (\kms) in the red hemisphere of SiO emission. Sky plane maps (upper panels) and dependence on position angle $\psi$ for  $0.5<R<1.5$ arcsec (central panels) and $1.5<R<2.5$ arcsec (lower panels) are shown for different $V_z$ intervals: $2<V_z<5$ \kms\ (left), $5<V_z<10$ \kms\ (middle) and the whole red hemisphere, $V_z>0$ (right). The curves are the results of sine wave fits listed in Table \ref{tab6}.}
\label{fig14}
\end{figure*}

Figure \ref{fig14} displays the map of $\langle V_z\rangle$ for the whole red-shifted hemisphere, $V_z>0$, as well as for two different $V_z$ intervals, $2<V_z<5$ \kms\ and $5<V_z<10$ \kms\ and two different $R$ intervals, $0.5<R<1.5$ arcsec and $1.5<R<2.5$ arcsec. Also shown, in each case, is the dependence of $\langle V_z\rangle$ on the position angle $\psi$. The results of sine wave fits of the form $\langle V_z\rangle=V_{z0}+V_{z1}\cos(\psi+\psi_0)$ are listed in Table \ref{tab6}. Note that for the red-shifted broad component in the whole $R$ range \citep{Hoai2019}, $\langle V_z\rangle \sim 4.7-0.34\cos(\psi-4^\circ)$ \kms. The values obtained here are very similar and are clearly dominated by expansion, leaving essentially no room for rotation. One might have expected significant rotation associated with the accretion disc responsible for collimating the jets, but such is not the case.

\subsection{A closer look at the polar jets}\label{sec7}

The presence of narrow polar jets in SiO emission and their absence from SO$_2$ emission argue against an interpretation in terms of a spherical shell ejected by star pulsation at short distances from the star, as described in \citet{Winters2003} and \citet{McDonald2016}.

The shell would be expected to be ejected at short radial distances from the star, at a scale corresponding to the escape velocity, meaning 4 to 6 au for a star mass of 1 to 2 solar masses.

The question would then arise of the distance over which the shell would slow-down to a velocity of 12 \kms\ or less. Gravity alone would imply a decrease inversely proportional to the square root of the distance, namely a radial velocity reaching 12 \kms\ at some 15 au. Indeed, as shown in the channel maps displayed in Figure \ref{figa7}, this distance could not significantly exceed some 20 au or so, excluding interpretations in terms of accelerated expansion of the spherical shell up to distances at the 100 au scale.

This is a region where SO$_2$ observations have given clear evidence for rotation, possibly combined with moderate expansion, both at the scale of a few \kms, typically 5. Such kinematic is at strong variance with that of a spherical shell expanding radially at velocities exceeding 12 \kms\ and reaching some 20 \kms.

Not only is the spherical shell interpretation irreconcilable with SO$_2$ observations, but it is unrelated to the observation of a dominant emission displaying clear bipolarity with a factor of 6 between polar and equatorial winds, for which a completely independent mechanism would have to be invoked.

\begin{table*}
  \centering
  \caption{Location of the jet projections on the sky plane for SiO emission.}
  \label{tab7}
  \begin{tabular}{cccc}
    \hline
    &&Mean (arcsec)&Rms(arcsec)\\
    \hline
    \multirow{2}{*}{Blue}&$x$&+0.044&0.146\\
    \smallskip
    &$y$&+0.076&0.150\\
    \multirow{2}{*}{Red}&$x$&+0.022&0.120\\
    &$y$&$-$0.028&0.110\\
    \hline
  \end{tabular}
  \end{table*}

\begin{figure*}
\centering
\includegraphics[width=0.7\textwidth,trim=0.cm 1.cm 0.cm 0.cm,clip]{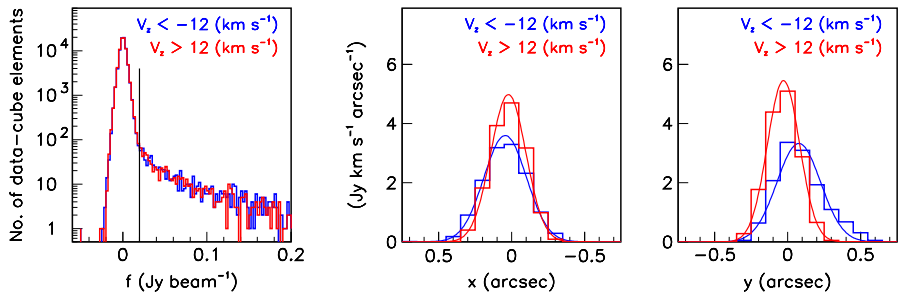}
\caption{SiO emission in the jets. Distributions of $f$ (left), $x$ (middle) and $y$ (right) are shown for $R<2.5$ arcsec and $V_z<-12$ \kms\ (blue) or $V_z>12$ \kms\ (red). In the rightmost panels a cut $f>0.02$ Jy beam$^{-1}$ is applied, as shown in the left panel.}
\label{fig19}
\end{figure*}

Figure \ref{fig19} (left) displays the noise distribution in the region of the jets, $R<2.5$ arcsec and $|V_z|>12$ \kms\ for SiO emission. A clean signal is seen above $f\sim0.02$ Jy beam$^{-1}$. Retaining flux densities exceeding this value, we show in the middle and right panels of Figure \ref{fig19} the $x$ and $y$ distributions of the blue-shifted and red-shifted jets separately. They are very clean with mean and rms values listed in Table \ref{tab7}. Taking as uncertainty on these measurements the rms deviation of the beam profile with respect to its mean gives differences between the red and blue values of respectively $\Delta x=-0.022\pm0.19$ arcsec and $\Delta y=-0.104\pm0.19$ arcsec; the larger value of $\Delta y$ is probably the result of the inclination of the star axis with respect to the line of sight: the average value of $|z|$ is $0.5\Delta y/\tan\varphi$, namely 0.4$\pm$0.4 arcsec for an inclination of 10\dego. In the case of $x$, the distribution of $\Delta x$ receives no contribution from the inclination of the jet axes and the rms deviation with respect to the mean provides an estimate of the opening angle of the jets: Rms($\Delta x$)/$|z|$, namely $\pm$9\dego\ for $|z| = 1$ arcsec, $\pm$18\dego\ for $|z| = 0.5$ arcsec.  Moreover, the numbers in Table \ref{tab7} show that the jets are launched within $\pm$25 au from the star, excluding a possible relation to a distant companion.

\section{CO emission}\label{sec6}

\subsection{$^{12}$CO(2-1) emission} \label{sec6.1}

\begin{figure*}
\centering
\includegraphics[width=0.8\textwidth,trim=0.cm 0.5cm 0.cm 0.cm,clip]{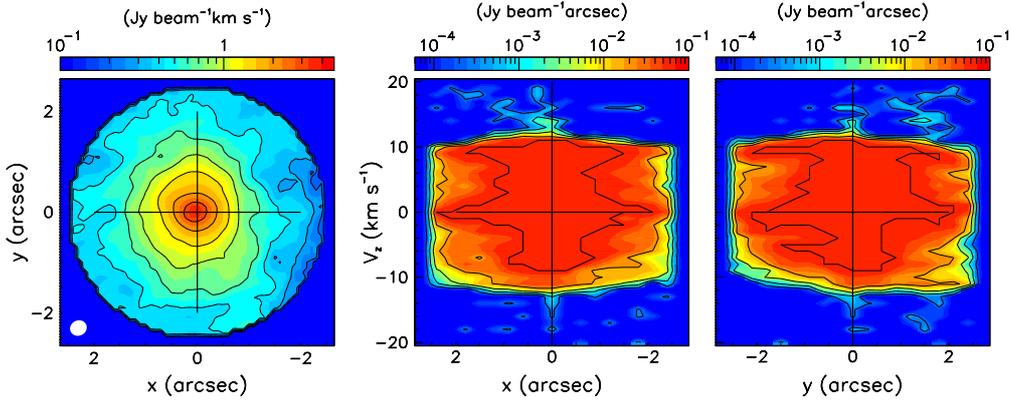}
\caption{Projections of the CO data-cube on ($x,y$) (left), ($x,V_z$) (middle) and ($y,V_z$) (right).}
\label{fig15}
\end{figure*}

\begin{figure*}
\centering
\includegraphics[width=0.75\textwidth,trim=0.cm .5cm 0.cm 0.cm,clip]{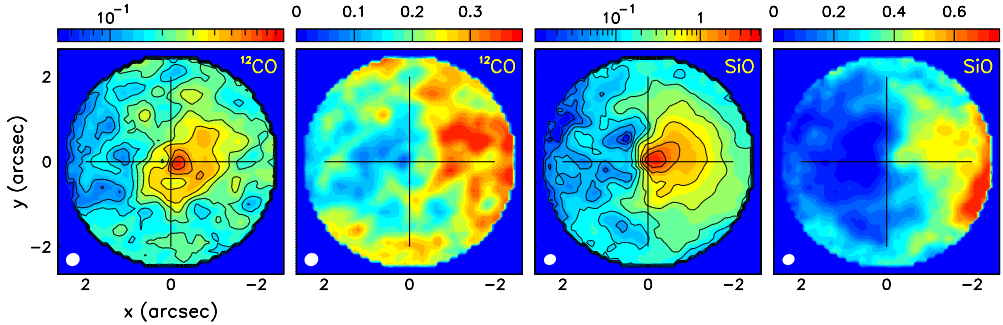}
\caption{Maps of the depletion of the data-cube are compared for CO (left panels) and SiO (right panels) emissions. In each pair of panels, the leftmost panel shows $f_{deplet}$ integrated over the full velocity range (Jy beam$^{-1}$\kms) and the rightmost panel shows the ratio $f_{deplet}/f_{uncut}$.}
\label{fig16}
\end{figure*}

A detailed study of CO emission, with emphasis on distances from the star in excess of $\sim$2 arcsec, was presented in \citet{Hoai2019}: it does not need to be repeated here. We limit instead the present study to a comparison with SiO observations, with the aim to contribute additional information to the description of the morpho-kinematics of the circumstellar envelope at short distances from the star ($r<2.5$ arcsec). Projections of the data-cube are displayed in Figure \ref{fig15} as was done for SiO emission in Figure \ref{fig9}. Comparing the two figures is very instructive. The presence of narrow polar jets at the limit of sensitivity is only revealed when applying a strong noise cut (4-$\sigma$ in Figure \ref{fig15}); additional clearer evidence is presented in Appendix \ref{appa1}. The P-V maps of Figure \ref{fig15} are red-blue asymmetric, as were those of SiO emission; but instead of displaying significantly different end points of the Doppler velocity spectra, they simply show that more gas has reached terminal velocity in the red hemisphere than in the blue. This was already clearly apparent on the global Doppler velocity spectrum displayed in Figure \ref{fig3}. Indeed, we know from \citet{Hoai2019} (their Figure 26) that absorption is twice as large on the blue side than on the red side for $|V_z|>8$ \kms\ and an independent confirmation is obtained from the study of $^{13}$CO emission presented in Section \ref{sec6.2} below. While the CO data-cube displays considerably less asymmetry than the SiO data-cube, striking similarity between the two is illustrated in Figure \ref{fig16} that compares maps of the depletion component $f_{deplet}$ in both absolute and relative terms. Here $f_{deplet}$, defined in Appendix \ref{appa3}, measures the missing emission associated with the depletion; namely the observed data-cube is written as $f=f_{uncut}-f_{deplet}$ where $f_{uncut}$ measures the emission of the centrally-symmetric intact data-cube, the observed data-cube $f$ being amputated by the contribution $f_{deplet}$ of the depletion. This figure provides remarkable evidence for the depletion to be present in both SiO and CO data in spite of the much smaller asymmetry that it produces in the latter: close inspection of the data-cube reveals its presence as a weak elliptical depression at $x=-0.4$, $-$0.5 and $-$0.6 arcsec.

Scanning through the data-cube in small steps of $x$ as was done for SiO in Figure \ref{fig13} shows that both SiO and CO maximal velocities are consistent with the same terminal velocity as obtained in \citet{Hoai2019} when assuming that SiO molecules explore a radial range reaching up to some 3 arcsec. It also shows that the jets are weak in CO emission, polar emission being enhanced at moderate velocities, as if they were slowing down at larger distances or if their aperture was broadening. Moreover, it shows that what was called the narrow component by \citet{Hoai2019}, equatorial emission at $V_z\sim0$ covering a broad range of $x$ and $y$, is absent from SiO emission, as if it were building up progressively and had no time to do so in the short radial range explored by SiO emission. Indeed, the star gravity, with an escape velocity of $\sim$4 \kms\ at $r=1$ arcsec and decreasing as $1/\sqrt{r}$, competes significantly against acceleration in the range explored by SiO, causing expansion to slow down as $r$ increases.

\subsection{$^{13}$CO(2-1) emission}\label{sec6.2}

\begin{figure*}
\centering
\includegraphics[width=0.75\textwidth,trim=0.cm .5cm 0.cm 0.cm,clip]{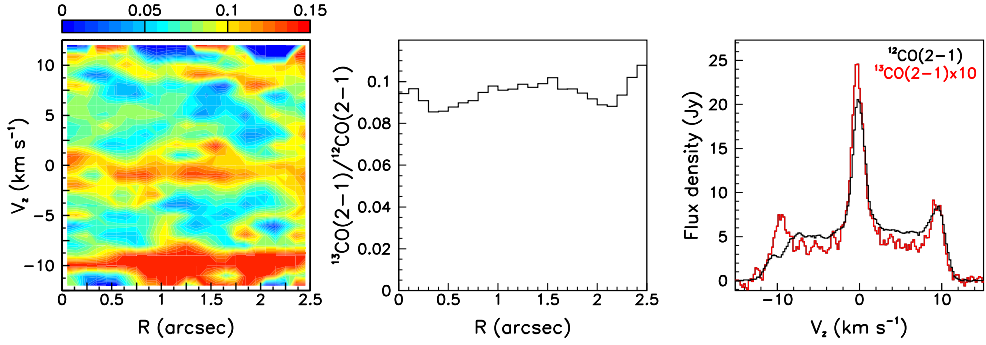}
\caption{Comparing $^{13}$CO and $^{12}$CO emissions ($R<2.5$ arcsec, corrected for beam sizes). Left: ratio $^{13}$CO/$^{12}$CO of the P-V maps. Middle: ratio of the $R$ distributions of the brightness integrated over Doppler velocity and position angle. Right: Doppler velocity spectra of $^{12}$CO emission (black) and  $^{13}$CO emission (red, scaled up by a factor 10).}
\label{fig18}
\end{figure*}

Figure \ref{fig18} compares $^{12}$CO(2-1) and $^{13}$CO(2-1) emissions at projected distances $R$ not exceeding 2.5 arcsec, the former being typically 10 times brighter than the latter. As had been remarked in Figure 20 of \citet{Hoai2019} the $R$ interval between 1 and 3 arcsec corresponds to the maximal enhancement of the high $|V_z|$ horns at the extremities of the Doppler velocity spectra, a result of the flux of matter being  maximal at intermediate stellar latitudes. The brightness ratio, averaged over position angle, is approximately $R$-independent but larger for Doppler velocities where absorption is more important, namely the narrow central component and the high $|V_z|$ horns, revealing the difference of optical thickness, $^{13}$CO emission being essentially optically thin. The effect is particularly important on the blue-shifted horn of the spectrum, causing it to disappear in the $^{12}$CO data.  Unfortunately, the relatively low value of the signal-to-noise ratio of $^{13}$CO observations prevents making a more refined analysis of the absorption.

Accounting for temperature, for an average absorption of 20$\pm$20\% of $^{12}$CO emission and for the values of the Einstein coefficients listed in Table \ref{tab2}, we measure a relative abundance ratio $^{12}$CO/$^{13}$CO of 9$\pm$2. This result corroborates the measurement by \citet{Cami2000} of CO$_2$ emission from distances corresponding to the region explored by SO$_2$ emission in the present work: they find a $^{12}$CO/$^{13}$CO ratio of the order of 10, however with a large uncertainty.

\section{Discussion and conclusion} \label{sec8}

\subsection{What has been learned: a summary}\label{sec8.1}

The results obtained in the present study at small distances from the star add up to those obtained  earlier at large distances from the star \citep{Hoai2019} to contribute an important number of new elements to our knowledge of the circumstellar envelope of EP Aqr. They draw complementary, but also significantly different pictures of its morphology and kinematics; at the same time as they help with a global understanding of the mechanisms at stake, they raise new question on the transition between the two regimes, such as: how do the jets disappear? how does the equatorial outflow (the narrow component) build up? How does the wind reach terminal velocity? It is useful, at this stage, to summarize what has been learnt.

1. Close to the star photosphere, at distances at the scale of 10 to 30 au, the kinematics is dominated by rotation, with a velocity of the order of $\sim$4 to 5 \kms, associated with a nascent radial expansion reaching a few \kms\ and displaying no significant anisotropy. The beam size, the line width and the sensitivity of the SO$_2$ observations prevent revealing more detailed features with sufficient confidence.

2. The line width of SO$_2$ emission is too large ($\sim$7.5 \kms\ FWHM) to be blamed exclusively on coherent Doppler broadening. It probably receives additional contribution from the turbulent regime that is expected to govern this region, host of shocks produced by the pulsation of the star and/or precursors of the nascent jets.

3. Two polar jets with a terminal velocity of some 20 \kms\ are launched from less than 25 au projected distance from the mass-losing star. Their opening angle is at the $\pm$10\dego\ to 15\dego\ scale and the measured splitting between their $y$ coordinates confirms the low value of the inclination angle of the star axis with respect to the line of sight. They are clearly seen in both SiO and CO emissions, however close to noise level in the latter case, but are absent from SO$_2$ emission. They are much weaker in CO than SiO emission, suggesting that they slow down and/or diverge at large distances from the star.

4. Observations are consistent with a same axis being the axis of rotation close to the star (SO$_2$), the jet axis (SiO and CO) and the axi-symmetry axis of the circumstellar envelope at distances in excess of $\sim$250 au (CO).  This makes it unnecessary to invoke different symmetry-breaking geometries at different distances from the star. To a precision of $\sim$10\dego\ the common axis projects on the sky plane $\sim$20\dego\ west of north and is inclined by $\sim$10\dego\ with respect to the line of sight.

5. The absence of detection of jet emission in the SO$_2$ data suggests that the jets acquire velocity over distances from the star between $\sim$20 au and $\sim$100 au, an interpretation that is consistent with the SiO and CO observations and agrees with observations of QX Pup \citep[]{SanchezContreras2018} that present some similarity with what is observed here.

6. The radial dependence of SO$_2$ emission is confined to the range where molecules are both excited and dissociated by the UV star light, below $\sim$30 au. SiO emission is observed to be confined to distances not exceeding $\sim$300 au, probably because of a combination of UV dissociation and the rapid aggregation of the gas onto dust grains. The boundary of the region that it populates is very sharp, again in agreement with observations of QX Pup \citep[]{SanchezContreras2018}. CO emission is slowly declining at larger distances, probably by dissociation from interstellar UV radiation.

7. A radial wind is building up at distances between $\sim$50 and $\sim$300 au from the star. A sensible description of the terminal velocity is given by the form $V\sim2+9\sin^2\alpha$ \kms\ with $\alpha$ being the stellar latitude, slightly larger in the red-shifted  than in the blue-shifted hemisphere.

8. The temperature is well described at distances in excess of $\sim$250 au by an exponential radial dependence of the form $\sim$109[K]$\exp$($-r$[arcsec]/3.1) and is maximal at intermediate latitudes. At smaller distances a form 106[K]/$r$[arcsec] is better adapted to extrapolation, implying temperatures of $\sim$500 to 600 K in the region explored by SO$_2$ emission.

9. A very strong blue-west/red-east asymmetry dominates the SiO data-cube. Evidence for its presence in the CO data-cube has been obtained, inducing qualitatively similar, but quantitatively much smaller asymmetry. In the SiO case, where terminal velocity has not yet been reached in the explored part of the blue-western quadrant, it causes the end points of the Doppler velocity spectra to differ in the blue-shifted and red-shifted hemispheres. The asymmetry is best described as a depletion of blue-western emission having sharp boundaries, starting near the star and expanding rapidly to the whole blue-western quadrant. Its presence in both SiO and CO data suggests that it is associated with a low gas density but its stronger effect on the SiO data-cube may be the result of a lower SiO/CO abundance ratio in the depletion or of an inefficient acceleration of the wind.

10. Absorption was evaluated by \citet{Hoai2019} at a typical level of 20$\pm$20\% from a comparison between $^{12}$CO(2-1) and $^{12}$CO(1-0) observations at large distances from the star. $^{13}$CO(2-1) emission confirms this result, however emphasising the importance of absorption for velocities smaller than $\sim -8$ \kms\ causing the blue-shifted horn of the velocity spectrum to disappear when observed in $^{12}$CO emission. While having little influence on the observation of the blue-western depletion at Doppler velocities larger than $\sim -8$ \kms, it weakens the significance of the blue-west/red-east asymmetry at smaller velocities.

11. The transition between rotation dominance close to the star and expansion dominance farther out occurs at distances from the star smaller than ~50 au, such that no significant rotation can be detected in the SiO data.

12. The CO line width is measured as 1.2 \kms\ (FWHM) in the equatorial region, leaving only little room for contributions such as turbulence and coherent Doppler broadening (flaring). On the contrary, both SO$_2$ and SiO line widths seem to receive significant contribution from turbulence, at the scale of a few \kms\ FWHM.

13. CO emission at large distances from the star reveals irregularities of the equatorial morpho-kinematics in the form of a spiral arc in brightness \citep{Homan2018} and of apparently uncorrelated concentric circles in expansion velocity. Both show a radial modulation with a period at the scale of 3 to 4 arcsec, meaning a time scale of 800 to 1200 years for a wind velocity of 2 \kms.

14. $^{13}$CO emission displays the same morpho-kinematics as $^{12}$CO emission but is optically thinner. The abundance ratio $^{12}$CO/$^{13}$CO is measured as 9$\pm$2.

15. While all above statements are the result of careful scrutiny of the properties of the relevant data-cubes, one must keep in mind the ambiguity and arbitrariness inherent to the under-determination of radio observations. The validity of many of these statements rests in part on a subjective appreciation of what we consider the most plausible physical interpretation of the observations being analysed.

\subsection{Constraints placed on possible interpretations}\label{sec8.2}

Lacking a convincing description of the morphology and kinematics of the circumstellar envelope of EP Aqr in terms of the physics and chemistry governing its dynamics, it is useful to  review the constraints that presently available observations and analyses can place on plausible interpretations and speculations.

The observation of narrow polar jets in the environment of a star in an early stage of evolution on the AGB was unexpected. While jets are common in astrophysics, their occurrence in stellar physics is normally restricted to young stellar objects or to post-AGB stars and pre-planetary or planetary nebulae. They are known to share universal features \citep{Pudritz2012}, in particular to be associated with accretion discs that surround the source and contribute to their collimation. In most cases, they refer to collimated gas flows having velocities an order of magnitude larger than that of the present jets (a few hundred \kms\ rather than 20 \kms). Yet, the terminology remains justified in the present case where their velocity is twice as large as the terminal wind velocity and an order of magnitude larger than that of the slow equatorial wind and where evidence has been obtained for collimation. But the difference between the present jets and fast jets observed in post-AGB stars must be kept in mind, together with what it implies in terms of the mechanism governing the dynamics.

The observation of a major asymmetry of the SiO data-cube, best described as a depletion of the blue-western quadrant, was also unexpected for a young AGB star. Breaking the spherical symmetry that governs the morpho-kinematics of red giants is normally discussed in the literature as occurring in the post-AGB phase with the launching of a super-wind. The main source of further symmetry breaking, this time beyond axi-symmetry rather than simple spherical symmetry, is considered to be binarity, which is widely recognized to play an important role in the evolution of mass-losing stars. A very abundant literature develops the above statements; here we can only quote a very few among the most recent, from which one can find one's way to a broader list of useful references: \citet{Soker2016}, \citet{Soker2017}, \citet{Sahai2018}, \citet{Lagadec2018}, \citet{Bollen2017}, \citet{Lykou2015}, \citet{PerezSanchez2013}.

To the extent that the jets are launched by stars, which they don't have to, binarity may suggest two different scenarios: one where the jets are launched by the mass-losing star \citep{Mastrodemos1999} and where the companion is simply focussing the wind that it is blowing toward the orbital plane where it produces spiral patterns associated with its wake; the other where the jets are launched by the companion \citep{Soker2000} and interact with the slow wind blown by the mass-losing star, producing shocks and depletions.
  
If the spiral observed in the equatorial plane \citep{Homan2018, Hoai2019} is to be interpreted as evidence for binarity, the inter-arm distance of some 350 to 400 au means a period of some 800 to 900 years at an expansion velocity of 2 \kms\ (the equatorial expansion velocity). This in turn implies, for a total mass of two solar masses a separation of some 100 to 120 au (note that \citet{Homan2018} use an equatorial expansion velocity of 12 \kms\ in their reasoning, resulting in a much shorter separation). In such a case, our observation that the jets are launched less than $\sim$25 au away from the central star seems to exclude that they be launched by the companion. Similarly, it seems to exclude that the tip of the depletion observed in the blue-western quadrant, at a distance not exceeding 40 au from the mass-losing star, be revealing the location of the companion. Therefore, if the spiral is taken as evidence for binarity the associated companion is probably unrelated to both the observed jets and the observed blue-western depletion. These need therefore to be interpreted independently from the equatorial spiral, possibly, but not necessarily as related to a closer companion.

The large ratio between polar and equatorial terminal wind velocities, a factor of $\sim$6,  suggests that the jets, or more precisely whatever mechanism is causing their acceleration, play a significant role in the acceleration of the slower wind. However, the impression that, qualitatively, the jets might accelerate the slow wind by entraining gas in their neighbourhood is not tenable quantitatively: they carry much too little momentum to support such an interpretation (we thank A. Zijlstra for clarification on this point).  In this context the case of a young  Herbig Ae star, HD 163296, \citep{Isella2016, Diep2019} is instructive: \citet{Klaassen2013} have shown that a collimated wind having a velocity just below 20 \kms observed in the proximity of the knots of a very high velocity jet (at the scale of $\sim$250 \kms) reveals an outflow from the accretion disc of the young star that has been simply heated up by the fast jet rather than being entrained by it as had been assumed earlier.\\

The jets are launched close to the star and reach quickly a velocity of 20 \kms\ while the rest of the wind builds up more slowly and reaches only $\sim$11 \kms. The jets have sharp boundaries and a specific identity, separate from that of the radial wind blowing at intermediate latitudes, which therefore cannot be simply described as the wings of the jets. The jets fade away at larger distances, suggesting that they interact with the ambient gas. The mechanism that is launching the jets must differ from the mechanism that accelerates the wind at sub-polar latitudes. The latter is barely able to compete against gravity near the equator.

While the present work has given evidence for narrow polar jets to be responsible for the higher velocity range of the observed SiO emission compared with that of CO emission, such difference is a general feature of the emission of dusty and low velocity outflows, as was noted earlier by \citet{Winters2003} and more recently by \citet{DeVicente2016}. It is often interpreted as resulting from star pulsations and is meant to be confined to the close neighbourhood of the star \citep{Winters2003,McDonald2016}. Recently, \citet{Decin2018} have observed the presence of wind velocities much larger than terminal in the inner regions of the circumstellar envelopes of oxygen-rich AGB stars IK Tau and R Dor and have discussed possible interpretations in terms of features other than narrow jets. The question then arises of how general, or how exceptional, is the presence of narrow polar jets in the nascent wind of AGB stars. The difficulty to detect such jets in geometries less favourable than that of EP Aqr makes it difficult to answer the question.

The presence of an important blue-western depletion is surprising. As explained by \citet{Soker2000} and illustrated by \citet{GarciaArredondo2004} such depletion can naturally occur as a result of the interaction of the jets with the slow wind blown by the mass-losing star. Similarities between the present observations of the nearby environment of EP Aqr and recent observations of that of QX Pup \citep{SanchezContreras2018} seem to favour this interpretation. However, there is no simple reason for it to happen on one jet and not on the other. We cannot think of any hint in the present or earlier data that might credibly suggest a cause of this red/blue asymmetry. In the case of QX Pup a similar asymmetry is observed, this time north/south rather than blue/red, the star axis making an angle of only $\sim$35\dego\ with the plane of the sky; the authors interpret it as caused by an early episode of violent mass loss, the star having ejected an important mass of gas and dust along its axis in one direction; we have no evidence for a similar effect in EP Aqr. 

The mechanism governing the launching of the observed jets remains therefore unknown. Evidence for rotation, with a velocity of $\sim$4 to 5 \kms\ at a radial distance of 10 to 30 au has been obtained; but such a rotation velocity, at a distance where the escape velocity is of the order of 10 \kms\ is a bit marginal to produce sufficient oblateness for jets to be launched by the mass-losing star. However, any mechanism that tends to inflate the equatorial region, rotation or else, may possibly generate sufficient pressure to push polar gas out along the axis and produce the observed polar structures (Zijlstra, private communication).

Another unanswered question concerns the observation of relatively stronger SiO than CO emission in the jets when compared with the surrounding gas: it may reveal an enhanced SiO/CO abundance ratio, but it may as well reveal different physical conditions in their environment; understanding what governs the SiO/CO abundance ratio in jets is a difficult question \citep[see for example]{Cabrit2012}.
  
Finally, we recall that the interaction of the circumstellar envelope with the interstellar medium is known to be important \citep{Cox2012, LeBertre2004} but simple geometric considerations exclude a possible relation to the observed blue-western depletion.

In summary the present observations and their analysis have given evidence for early symmetry breaking of the morpho-kinematics of the circumstellar envelope of an AGB star, in strong contrast with the commonly accepted idea that such symmetry breaking normally occurs at the end of the AGB era, before the planetary nebula phase. Two narrow polar jets have been detected and whatever mechanism is causing their acceleration is likely playing a role in the acceleration of the slower wind at sub-polar latitudes. Evidence has been given for a strong depletion of the blue-western quadrant in both SiO and CO emissions, however much weaker in the latter case than in the former. We have shown that if the observed equatorial spiral is taken as evidence for binarity the associated companion is probably unrelated to both the observed jets and the observed blue-western depletion. These need therefore to be interpreted independently from the equatorial spiral.

Many questions remain unanswered: which is the mechanism that governs the launching of the jets? which is the mechanism that governs the depletion of the blue-western quadrant? which is the mechanism that governs the acceleration of the wind to terminal velocity? what causes the observed asymmetry between the blue-shifted and red-shifted hemispheres? what precisely causes the difference between CO and SiO emission in the jets? how does the equatorial disc (the narrow component) build up? how do the jets slow down and/or open up? what causes the modulations observed in the equatorial plane at large distances (a spiral of intensity and circles of velocity)?
  
These observations illustrate the complexity of the morpho-kinematics of nascent winds and warn against too hasty interpretations in the absence of observations of sufficient spatial and spectral resolutions. It might well be that other oxygen-rich AGB stars, such, for example, as RS Cnc \citep{Hoai2014, Nhung2015}, would reveal similar complexity and early deviation from spherical symmetry when observed in greater detail: the seeds of the strong distortions that govern the formation of planetary nebulae may be present at an earlier stage of the star evolution than commonly assumed. The need for further detailed observations of other oxygen-rich AGB stars is made evident.

\section*{ACKNOWLEDGEMENTS}

We express our deep gratitude to the referee, Professor Albert Zijlstra, for many pertinent comments that helped greatly with improving the quality of the manuscript. This paper makes use of the following ALMA data: 2016.1.00026.S and 2016.1.00057.S. ALMA is a partnership of ESO (representing its member states), NSF (USA) and NINS (Japan), together with NRC (Canada), NSC and ASIAA (Taiwan), and KASI (Republic of Korea), in cooperation with the Republic of Chile. The Joint ALMA Observatory is operated by ESO, AUI/NRAO and NAOJ. This work was supported by the Programme National Physique et Chimie du Milieu Interstellaire (PCMI) of CNRS/INSU with INC/INP co-funded by CEA and CNES. The Hanoi team acknowledges financial support from VNSC/VAST, the World Laboratory, the Odon Vallet Foundation and the Rencontres du Viet Nam. This research is funded by Vietnam National Foundation for Science and Technology Development (NAFOSTED) under grant number 103.99-2018.325.

%%%%%%%%%%%%%%%%%%%%%%%%%%%%%%%%%%%%%%%%%%%%%%%%%%



%\bibliographystyle{mnras}


%%%%%%%%%%%%%%%%%%%%%%%%%%%%%%%%%%%%%%%%%%%%%%%%%%

%%%%%%%%%%%%%%%%% APPENDICES %%%%%%%%%%%%%%%%%%%%%

%%%%%%%%%%%%%%%%%%%%%%%%%%%%%%%%%%%%%%%%%%%%%%%%%%

\appendix
\section{}
We collect here additional material that, although not essential, strengthens the arguments developed in the main text.

\subsection{The jets in CO and SiO emissions}\label{appa1}

\begin{figure}
\centering
\includegraphics[width=0.45\textwidth,trim=0.cm 0.cm 0.cm 0.cm,clip]{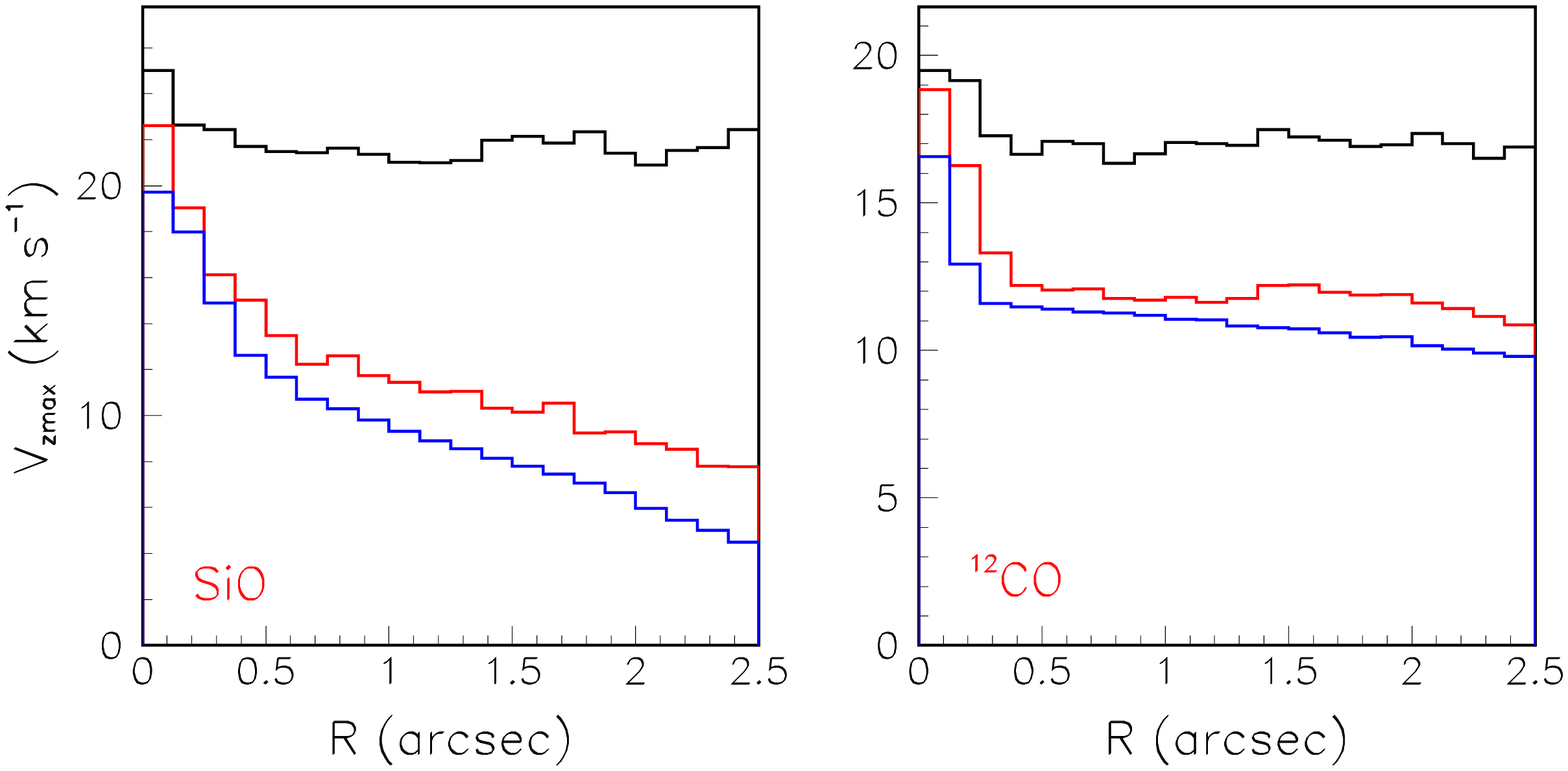}
\caption{Dependence on $R$ of the end points of the centrally-symmetrized Doppler velocity spectra ($V_{zmax}=-V_{zmin}$) for SiO (left) and CO (right) observations. In each panel cuts of 1 (black), 2 (red) and 3 (blue) noise $\sigma$'s have been applied.}
\label{figa1}
\end{figure}

\begin{figure}
  \centering
  \includegraphics[width=0.47\textwidth,trim=0.cm .5cm 0.cm 0.cm,clip]{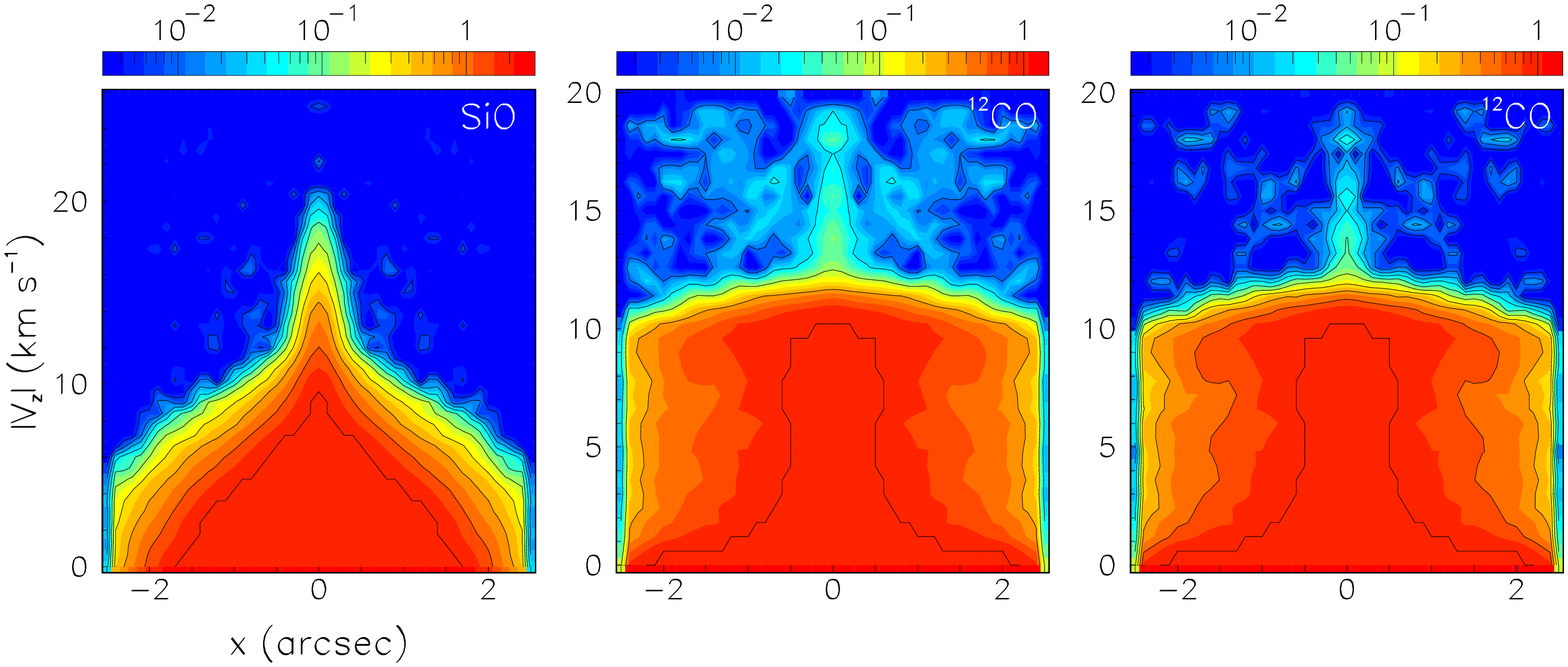}
  \caption{P-V maps of the intensity (Jy arcsec$^{-1}$) for centrally-symmetrized SiO (left panel, 2-$\sigma$ noise cut) and CO (right panels) observations. In the CO case the middle panel is for a 1.5-$\sigma$ noise cut, the right panel for a 2-$\sigma$ cut.}
  \label{figa2}
\end{figure}

\begin{figure}
\centering
\includegraphics[width=0.45\textwidth,trim=0.cm 0.cm 0.cm 0.cm,clip]{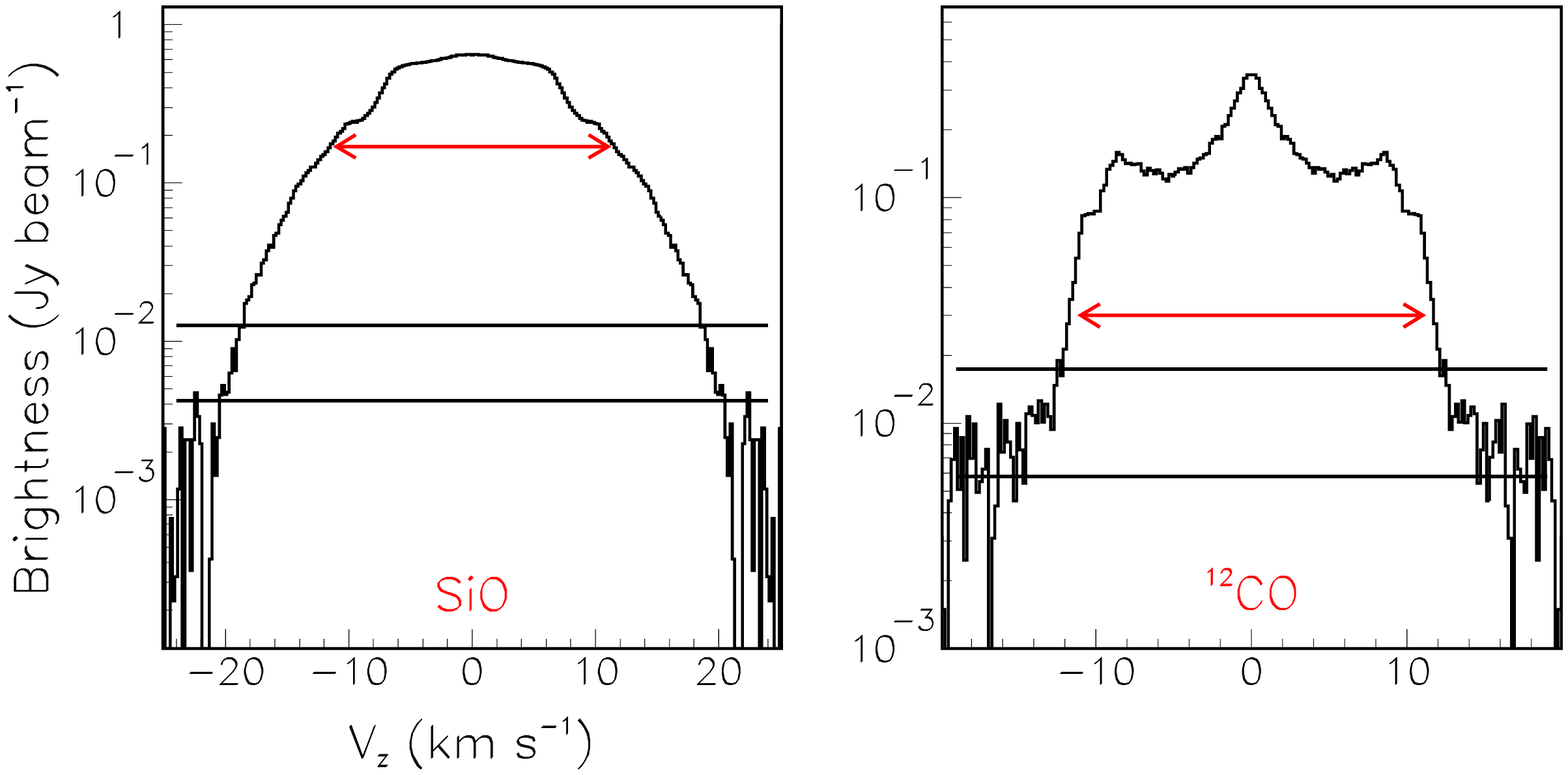}
\caption{Doppler velocity spectra of SiO (left) and CO(right) observations in the polar region $R<0.25$ arcsec in semi-logarithmic coordinates. In both panels, the lines show the 1-$\sigma$ and 3-$\sigma$ noise levels. The arrows show the ranges covered by the slower wind.}
\label{figa3}
\end{figure}

Figure \ref{figa1} compares the $R$-distributions of the end points, $V_{zmax}=-V_{zmin}$ , of the Doppler velocity spectra for SiO and CO observations respectively, using centrally symmetrized data-cubes $f_s$ as defined at the end of Section \ref{sec3}. The polar jets are now seen in both SiO and CO emissions as steep enhancements near the origin and Figures \ref{figa2} and \ref{figa3} show that they are close to noise in each case, particularly in the latter. Jets are usually considered as evidence for the presence of a companion accreting gas and an abundant literature describes the mechanisms at stake in their production \citep[][and references therein]{Soker2016}; other possible formation mechanisms are from magnetic fields \citep{Vlemmings2006} or as an effect of star rotation \citep{GarciaSegura1999, GarciaSegura2014}. Such mechanisms would naively be expected to produce similar effects in each of the red and blue hemispheres, which is not the case here.

\subsection{Asymmetry of the SiO and CO data-cubes}\label{appa2}

\begin{figure}
\centering
\includegraphics[width=0.45\textwidth,trim=0.cm 0.cm 0.cm 0.cm,clip]{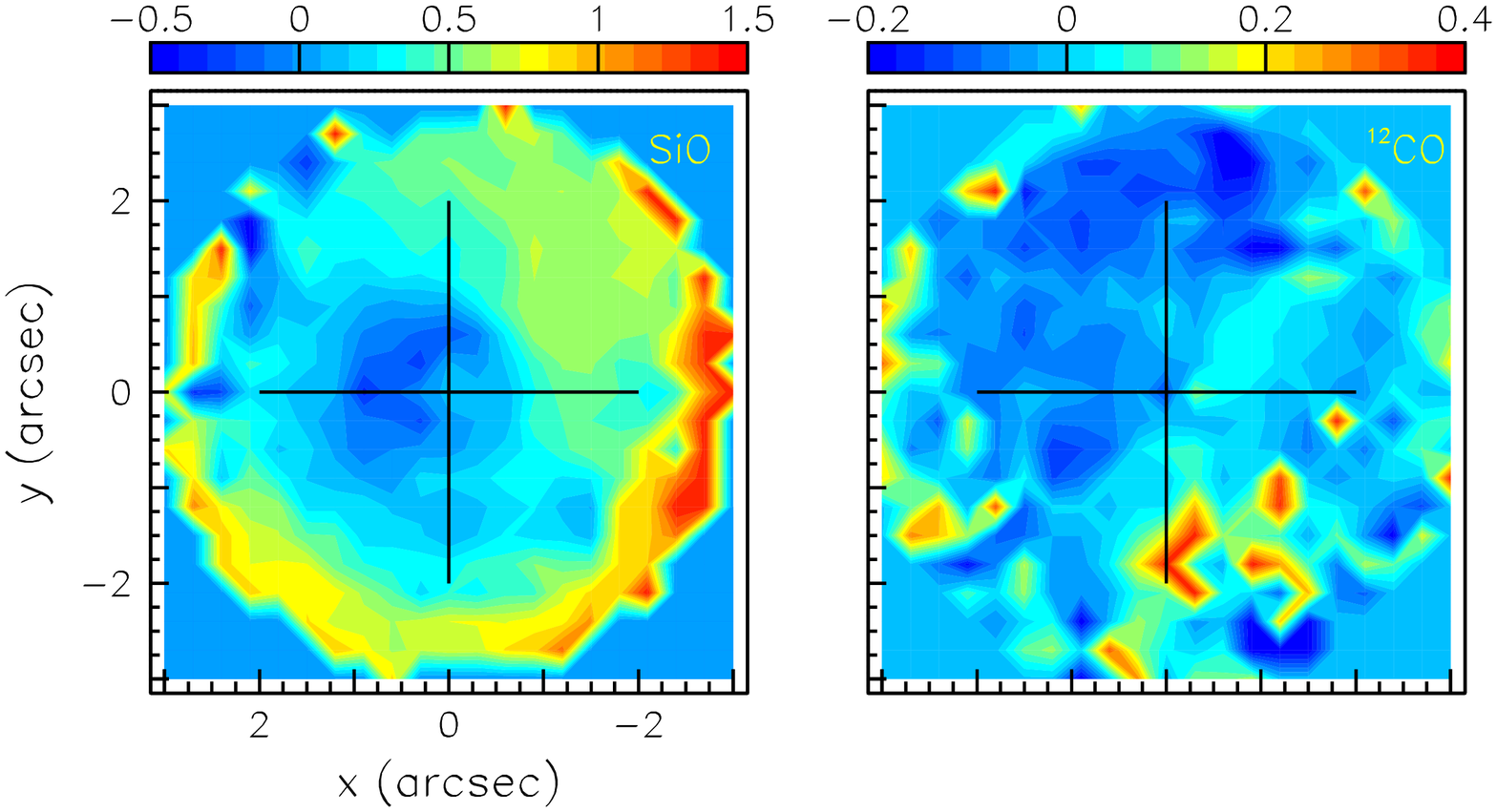}
\caption{Maps of $\Delta V_z(x,y)=[V_{zmax}(-x,-y)+V_{zmin}(x,y)]/V_{zmax}(-x,-y)$ using a 3-$\sigma$ noise cut are compared for SiO (left) and CO(right) emissions. Note the different colour scales.}
\label{figa4}
\end{figure}

Figure \ref{figa4} compares the SiO and CO maps of the quantity $\Delta V_z(x,y)=[V_{zmax}(-x,-y)+V_{zmin}(x,y)]/V_{zmax}(-x,-y)$, that would cancel if the data-cubes were centrally symmetric.  

\begin{figure*}
\centering
\includegraphics[width=0.75\textwidth,trim=0.cm 0.cm 0.cm 0.cm,clip]{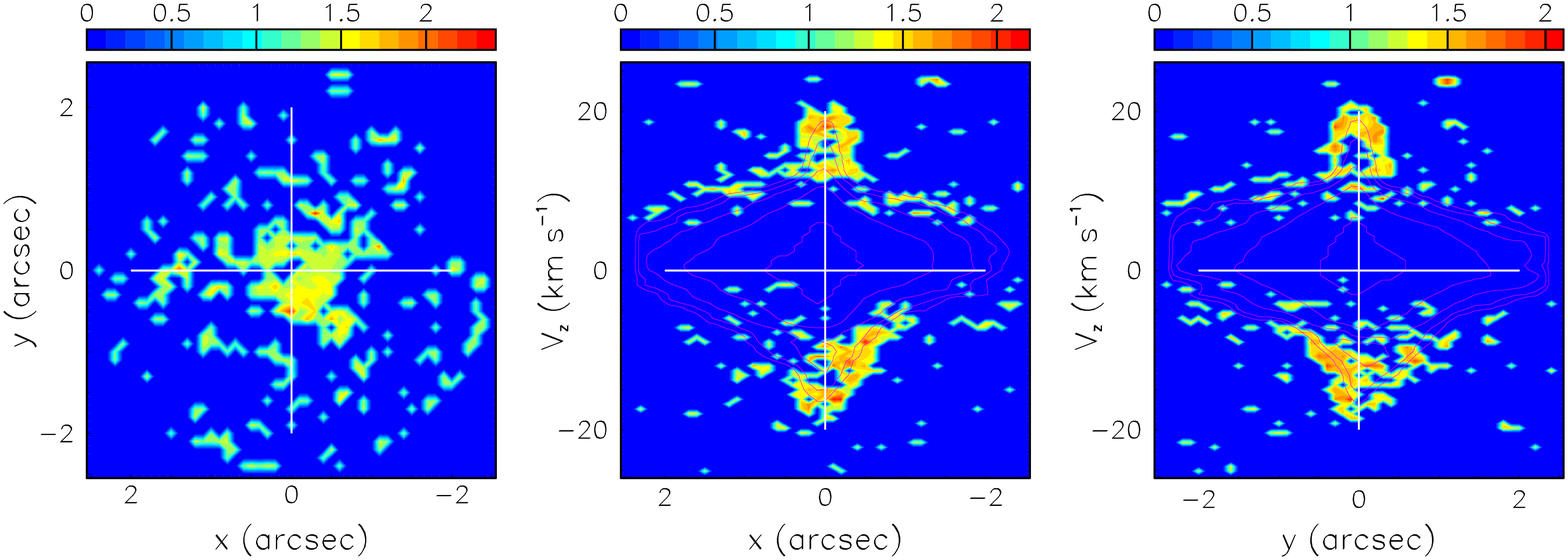}
\caption{Mapping of the high gradient regions of the SiO data-cube in $(x,y)$ (left), $(x,V_z)$ and $(y,V_z)$ (right). The colour scale measures the ratio Rms($f$)/Mean($f$) calculated over the surrounding $3\times3\times3$ data-cube elements. The contours displayed in the rightmost $(x,V_z)$ and $(y,V_z)$ panels are from Figure \ref{fig9} and depict the limits of significant emission.}
\label{figa5}
\end{figure*}

To further illustrate the sharpness of the boundaries of the depletion observed in the SiO data-cube, we trace the regions of high gradient defined as large values of the ratio $\gamma=$Rms($f$)/Mean($f$) calculated over a $3\times3\times3$ (3$\times$3 pix$^2$ and 3 velocity bins) small volume around each point. The result, displayed in Figure \ref{figa5} for $\gamma>1.4$, reveals enhanced gradients in the jet and depletion regions. 

\subsection{A closer look at the SiO data-cube}\label{appa3}

In Section \ref{sec5.3} we commented on the bias caused by the blue-western depletion of the SiO data-cube when attempting a distinction between rotation and expansion. We illustrate this in the present section by defining centrally symmetric and depletion data-cubes, respectively $f_{uncut}$ and $f_{deplet}$ such that the observed data-cube is the difference between the two: $f=f_{uncut}-f_{deplet}$. Precisely we associate to each pair of centrally opposite data-cube elements $f(x,y,V_z)$ and $f(-x,-y,-V_z)$ the maximal and minimal values of the two: $f_{max}$ and $f_{min}$ and define $f_{uncut}$ as $f_{max}$ at both $(x,y,V_z)$ and $(-x,-y,-V_z)$ and $f_{deplet}$ as $f_{max}-f(x,y,V_z)$ at $(x,y,V_z)$ and as $f_{max}-f(-x,-y,-V_z)$ at $(-x,-y,-V_z)$, one of which cancels by construction.

\begin{figure*}
\centering
\includegraphics[width=0.8\textwidth,trim=0.cm 0.cm 0.cm 0.cm,clip]{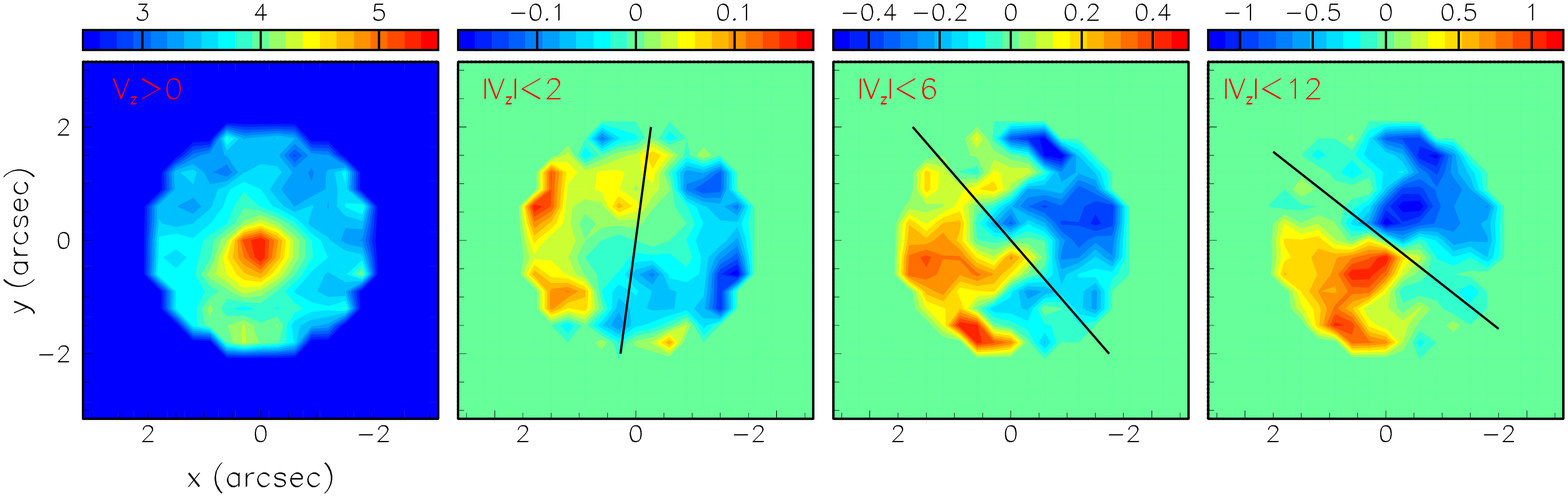}
\caption{SiO observations. Left: map of the mean Doppler velocity (\kms) in the red hemisphere of the centrally-symmetrized data-cube $f_{uncut}$. Rightmost panels: maps of the mean Doppler velocity of $f_{uncut}$ for $|V_z|$ <2 \kms, <6 \kms\ and <12 \kms\ from left to right. The lines indicate the axes of anti-symmetry.}
\label{figa6}
\end{figure*}

Figure \ref{figa6} displays the centrally-symmetrized map of $\langle V_z \rangle$ in the blue hemisphere (by definition, the map in the red hemisphere is obtained from it by central symmetry); it is similar (up to a sign) to the red hemisphere map shown in Figure \ref{fig14}, the red hemisphere being essentially unaffected by the depletion. Also shown in Figure \ref{figa6} are centrally-symmetrized maps of $\langle V_z \rangle$ for three intervals of $|V_z|$, <2 \kms, <6  \kms\ and <12  \kms. It is tempting to infer from these maps a strong dominance of rotation. But in reality, the effect is mostly the result of the bias introduced by the blue-western depletion; it causes $f_{uncut}$ to be dominated by the red-eastern quadrant, namely to produce maps that tend to be red-shifted in the eastern hemisphere and blue-shifted in the western hemisphere. It results in an apparent rotation opposite to that observed for SO$_2$ emission and illustrates the difficulty to make a reliable distinction between rotation and expansion in the presence of a strong blue-west/red-east asymmetry as is the case here.

\begin{figure*}
\centering
\includegraphics[width=0.75\textwidth,trim=0.cm 1.cm 0.cm 0.cm,clip]{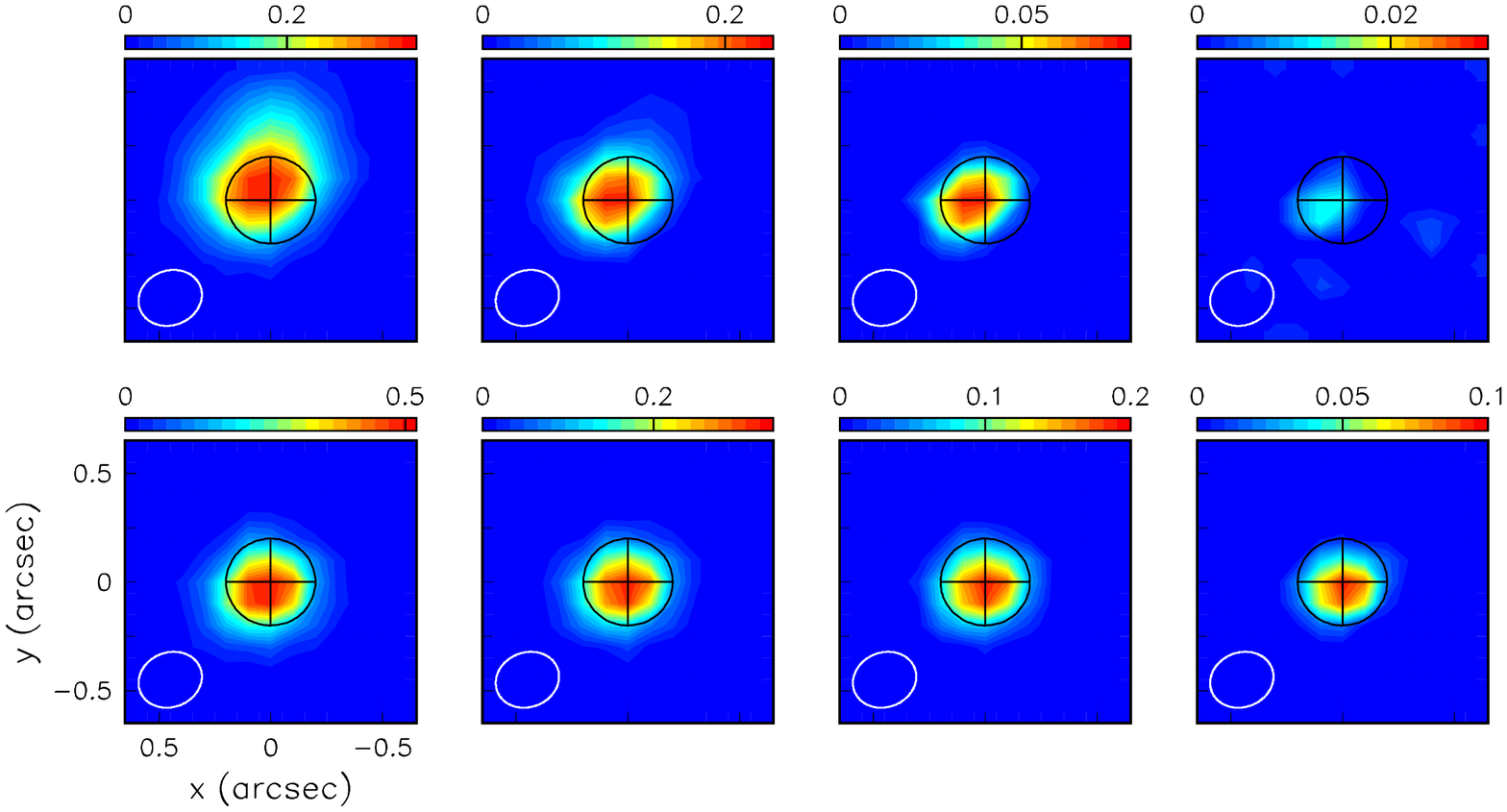}
\caption{Channel maps of SiO emission in the jet region using a 0.01 Jy beam$^{-1}$ noise cut. Doppler velocity intervals run from left to right for $|V_z|$ between 12 and 14, 14 and 16, 16 and 18 and 18 and 20 \kms\ respectively. The upper row is for the blue-shifted jet, the lower row for the red-shifted jet. The colour scales are in units of Jy beam$^{-1}$\kms.}
\label{figa7}
\end{figure*}

% Don't change these lines
\bsp	% typesetting comment
\label{lastpage}
\end{document}